\documentclass[superscriptaddress,noshowpacs,noshowkeys, twocolumn, floatfix,aps, prb,reprint,citeautoscript]{revtex4-1}
\usepackage[english]{babel}
\usepackage{fontenc}
\usepackage{graphicx}
\usepackage{amsmath}		
\usepackage{epstopdf}
\usepackage{amssymb}
\usepackage{amsbsy}
\usepackage{amscd}
\usepackage{color}
\usepackage{bbm}
\usepackage{braket}
\usepackage{bm}
\usepackage{siunitx}
\usepackage{placeins}
\usepackage[normalem]{ulem} 
\usepackage{verbatim}
\usepackage{url}
\usepackage[verbose,hypertexnames=false,bookmarksopenlevel=1,filecolor=blue,
linkcolor=blue,citecolor=blue,urlcolor=blue,pdfstartview=FitH,bookmarksopen,bookmarksnumbered,
colorlinks,plainpages=false,linktocpage]{hyperref}
\graphicspath{{./figs/}}
\begin{document}
\renewcommand{\figurename}{Fig.}
\newcommand{\mb}[1]{\mathbf{ #1 }}
\newcommand{\mbs}[1]{\boldsymbol{ #1 }}
\newcommand{\mc}[1]{\mathcal{ #1 }}
\newcommand{\mcb}[1]{\boldsymbol{\mathcal{ #1}}}

\newcommand{\pauli}{\boldsymbol{\sigma}}
\newcommand{\valley}{\boldsymbol{\tau}}

\newcommand{\red}[1]{\textcolor{red}{#1}}
\newcommand{\blue}[1]{\textcolor{blue}{#1}}
\newcommand{\green}[1]{\textcolor{green2}{#1}}
 \definecolor{green2}{RGB}{25, 158, 45}

\newcommand{\sdo}{\tilde{\Lambda}_\text{sd}} 
\title{Reliable modeling of weak antilocalization for accurate spin-lifetime extraction}

\author{Michael Kammermeier}
\email{michael.kammermeier@vuw.ac.nz}
\affiliation{School of Chemical and Physical Sciences and MacDiarmid Institute for Advanced Materials and Nanotechnology, Victoria University of Wellington, P.O. Box 600, Wellington 6140, New Zealand}
\affiliation{Center for Science and Innovation in Spintronics (Core Research Cluster), Tohoku University, Sendai 980--8577, Japan}
\author{Takahito Saito}
\affiliation{Department of Materials Science, Tohoku University, Sendai 980--8579, Japan}
\author{Daisuke Iizasa}
\affiliation{Department of Materials Science, Tohoku University, Sendai 980--8579, Japan}
\author{Ulrich Z\"ulicke}
\affiliation{School of Chemical and Physical Sciences and MacDiarmid Institute for Advanced Materials and Nanotechnology, Victoria University of Wellington, P.O. Box 600, Wellington 6140, New Zealand}
\author{Makoto Kohda}
\affiliation{Center for Science and Innovation in Spintronics (Core Research Cluster), Tohoku University, Sendai 980--8577, Japan}
\affiliation{Department of Materials Science, Tohoku University, Sendai 980--8579, Japan}
\affiliation{Center for Spintronics Research Network, Tohoku University, Sendai 980--8577, Japan}
\affiliation{Division for the Establishment of Frontier Sciences, Tohoku University, Sendai 980--8577, Japan}

\date{\today }
\begin{abstract}
We examine models for the quantum-interference correction to the magneto-conductivity in two-dimensional electron gases  (2DEGs) with both Rashba and Dresselhaus spin-orbit coupling for their applicability to experimental data fitting.
In particular, we compare the Landau-quantized Cooperon approach, which is mostly only numerically treatable, and the quasi-classical approximation that was recently employed to obtain an explicit solution for arbitrary Rashba and Dresselhaus spin-orbit couplings [Marinescu~\textit{et al.}, \hyperlink{https://doi.org/10.1103/PhysRevLett.122.156601}{Phys. Rev. Lett. {\bf 112}, 156601 (2019)}].
It is found that the quasi-classical approximation yields significantly different results even to lowest order in the magnetic field and appears unsuitable for reliable parameter fitting.
The discrepancy emerges when a sum over Landau levels is replaced by an integral over wave vectors.
Substantial improvement is achieved by supplementing the quasi-classical model with the first two corrections given by the Euler-MacLaurin formula for approximating a sum by an integral. 
Corresponding modifications are, however, only feasible  in special spin-orbit parameter configurations where the mixing of Landau bands is negligible 
and a closed-form solution that accounts for Landau quantization is also available.
Such a scenario appears in a parameter regime where a persistent spin helix emerges and a transition between weak localization and weak antilocalization takes place.
Combining recent findings, we derive a generalized closed-form expression for the magneto-conductivity correction applicable to generic 2DEGs that are grown along a crystal direction with at least two growth-direction Miller indices equal in modulus. 
The result is a function of spin lifetimes of the long-lived spin textures and valid close to the persistent-spin-helix regime.
The accuracy of the derived formula is validated by comparing with results from numerical diagonalization of the multiband Cooperon as well as a recently established Monte-Carlo-based real-space simulation in exemplary (001)-, (113)-, and (110)-2DEGs.
\end{abstract}

\maketitle
%
%
\allowdisplaybreaks
\section{Introduction}

In 1980, Hikami~\textit{et al.} discussed the importance of spin-orbit (SO) coupling  for the  quantum-interference correction to the conductivity. \cite{Hikami1980}
In diffusive conductors without SO coupling, the constructive interference of counter-propagating electrons in closed-loop scattering paths leads to an enhanced back-scattering probability and concomitantly a reduction of the conductivity, known as weak localization.\cite{Gorkov1979,Anderson1979}
This effect, however, can be reversed by SO-induced spin decoherence, which is therefore called weak antilocalization.
Distinctive features of weak antilocalization appear as a low-field negative magneto-conductivity with local extrema, related to the competition of spin relaxation and magnetic dephasing.
Fitting experimental data from magneto-conductance measurements with theoretical expressions has to date evolved into a useful tool for extraction of SO coupling strength, spin lifetime, and other diffusion and transport parameters and can be employed to various kinds of semiconductor quantum structures in diverse mesoscopic configurations.~\cite{Knap1996,Hassenkam1997,Zumbuehl2002,Schapers2006,Thillosen2006,
Guzenko2007,Kunihashi2009,Caviglia2010,Moriya2014,
Weperen2015,Kammermeier2016,Kammermeier2017,Jespersen2018}
Since the fitting procedure needs to incorporate a rather extensive parameter space, the theoretical result should ideally be available in analytic or closed form.

While Hikami~\textit{et al.} considered spin-flip scattering at impurities  as the central mechanism for spin relaxation (Elliott-Yafet mechanism),\cite{elliott,yafet} in diffusive conductors with broken inversion symmetry the D'yakonov-Perel'\cite{perel} spin relaxation due to a randomization of spin precession is often more relevant.
In III-V semiconductors with zinc-blende structure, the broken inversion symmetry becomes manifest in  wave-vector ($\mb{k}$) dependent spin splittings due to $k$-linear Rashba~\cite{Rashba1960,Bychkov1984} and $k$-linear and $k$-cubic Dresselhaus~\cite{Dresselhaus1955} SO coupling.
Computing the magneto-conductivity correction with generic Rashba and Dresselhaus SO couplings is in general only achieved numerically.
Technically, such a calculation requires a summation over the eigenvalues of the multiband Cooperon propagator, which characterizes the closed-loop quantum interference.\cite{Pikus1995}
Since $k$-linear SO terms non-trivially mix spin components and adjacent Landau levels, the Cooperon eigenvalues can be obtained analytically only in a few specific cases as summarized below.

(i)~In bulk semiconductors with pure $k$-cubic Dresselhaus SO coupling, Altshuler~\textit{et al.} derived a closed-form expression for the magneto-conductivity correction.\cite{Altshuler1981b}
(ii)~2DEGs with [001] growth direction were first studied by Iordanskii~\textit{et al.}, who took into account $k$-linear and $k$-cubic Dresselhaus SO coupling and derived an analytic formula for the magneto-conductivity correction.\cite{Iordanskii1994}
At the same time, the result is also valid for pure Rashba SO coupling because the Hamiltonians for Rashba and $k$-linear Dresselhaus SO coupling can be converted into each other by a unitary transformation.\cite{Pikus1995,Knap1996}
(iii)~Another special case was pointed out by Pikus~\textit{et al.}\cite{Pikus1995} if in a (001)-2DEG the Rashba SO coefficient equals the $k$-linear Dresselhaus SO coefficient and the $k$-cubic Dresselhaus SO coupling is neglected.
This observation was later adopted to (110)-2DEGs without Rashba SO coupling, which has analogous properties.\cite{Hassenkam1997}
Such special scenarios conform to the formation of a persistent-spin-helix (PSH) symmetry, enabling homogeneous and helical persistent spin textures.\cite{Schliemann2003,Bernevig2006,Schliemann2017,Kohda2017} 
These spin textures exist in 2DEGs with at least two growth-direction Miller indices equal in modulus if Rashba and $k$-linear Dresselhaus SO  coefficients are suitably matched and the $k$-cubic Dresselhaus SO coupling is negligible.\cite{Kammermeier2016PRL}

Recently, there have been several attempts to bridge the gap between these particular solutions and cover a wider SO parameter regime. 
To avoid the complications arising from the coupled Landau levels, an attractive approach is to assume a quasi-classical approximation, in which the magnetic field leads only to an additional dephasing term  and the energy spectrum remains quasi-continuous. 
Based on this approach, Marinescu~\textit{et al.} presented an approximate expression for the magneto-conductivity correction for arbitrary Rashba and Dresselhaus SO couplings in a (001)-2DEG.\cite{Marinescu2019} 
Other works employed a perturbative expansion of the Cooperon eigenvalues based on the exact solvability at the PSH-symmetry point.\cite{Kammermeier2016PRL,Weigele2020} 
Deviations from this symmetry can induce a crossover from weak localization to weak antilocalization, which is accompanied by significant changes in the magneto-conductivity.\cite{Kohda2012}
The PSH-symmetry breaking arises from the presence of $k$-cubic Dresselhaus SO coupling and small changes of the ratio of $k$-linear SO coefficients and limits the lifetime of the formerly persistent spin textures as studied in detail for 2DEGs with general growth direction in Refs.~\onlinecite{Kammermeier2016PRL,Iizasa2020}.
Kammermeier~\textit{et al.}~\cite{Kammermeier2016PRL} provided a general closed-form expression near the PSH symmetry point for arbitrarily-oriented 2DEGs with $k$-linear SO coupling utilizing the quasi-classical treatment of the magnetic field.
Weigele~\textit{et al.}~\cite{Weigele2020} derived an analogous solution for a (001)-2DEG but also accounting for $k$-cubic Dresselhaus SO coupling and Landau quantization.

In this paper, we critically examine the matter of applicability of these new models for accurate experimental parameter fitting.
Although qualitatively the magneto-conductivity correction obtained from the quasi-classical approximation and the exact consideration of the Landau quantization exhibit  the same trends, we find significant quantitative discrepancies.
Remarkably, the deviations are most prominent for low magnetic fields, where the magnetic dephasing time is much longer than the inelastic scattering time.
To lowest order, the magneto-conductivity correction scales linearly with the magnetic field in the quasi-classical approximation while it scales parabolically in the Landau-level picture.
The origin of this discrepancy lies in the inaccuracy that occurs when replacing a sum over Landau levels by an integral over wave vectors. 
The error can be strongly diminished by including in the quasi-classical model the first two corrections given by the Euler-MacLaurin formula for integral approximation of a sum.
Not only does this allow to recover the correct quadratic scaling of the magneto-conductivity in the low-field limit, it also yields generally excellent agreement with the Landau-level approach.
Unfortunately, the applicability of the Euler-MacLaurin formula  is limited to situations when the Landau levels decouple and the Cooperon spectrum is purely quadratic in the wave vector, which does not hold for general SO coupling and is inapplicable to correct the quasi-classical model of Marinescu~\textit{et al}.\cite{Marinescu2019}, derived for arbitrary Rashba and Dresselhaus SO couplings.

On the other hand, the model of Weigele~\textit{et al.}\cite{Weigele2020} that approximates the Cooperon spectrum near the PSH-symmetry but includes Landau quantization displays high accuracy in a broad parameter range.
The underlying reason is that in this regime a gauge transformation allows to decouple the Landau levels and the Cooperon spectrum resembles that of a system without $k$-linear SO coupling, which can be treated then analogously to case (i) of Altshuler~\textit{et al.}\cite{Altshuler1981b}
Combining the recent findings of Kammermeier~\textit{et al.}~\cite{Kammermeier2016PRL} and Iizasa~\textit{et al.}~\cite{Iizasa2020}, we can generalize the model of Weigele~\textit{et al.}~\cite{Weigele2020} to arbitrarily oriented 2DEGs in the vicinity of the PSH regime.
The derived magneto-conductivity correction is given as a function of the finite lifetimes of the long-lived spin textures, resulting from $k$-cubic Dresselhaus SO coupling and small changes of the optimal ratio of $k$-linear SO  coefficients. 
We demonstrate the applicability of the obtained closed-form expression explicitly by a comparison with numerical diagonalization of the multiband Cooperon and a Monte-Carlo-based real-space simulation as developed by Sawada~\textit{et al.}\cite{Sawada2017}
for 2DEGs grown along [001], [113], and [110] crystal directions.
The exemplary (001)-2DEG corresponds to the GaAs/AlGaAs quantum-well  structure designed in Ref.~\onlinecite{Saito2021}, where the closed-form expression was used for all-electrical evaluation of the spin lifetimes.
The extracted values were in excellent agreement with results independently obtained by exploring the weak-localization anisotropy in wire geometries with in-plane magnetic field.

This paper is organized as follows.
In Sec.~\ref{sec:basics}, we introduce the low-energy Hamiltonian for a generic 2DEG with at least two growth-direction Miller indices equal in modulus, the arising long-lived spin textures as well as the weak (anti)localization correction and its modifications in presence of an out-of-plane magnetic field.
The magneto-conductivity correction in a system without spin-splitting is studied in Sec.~\ref{sec:noSOC}, while Sec.~\ref{sec:SOC001} focuses on (001)-2DEGs with SO coupling.
In Sec.~\ref{sec:SOCaab}, we give an expression for the magneto-conductivity correction for general 2DEGs near the PSH symmetry point and discuss its applicability using several examples in comparison with former models.

\section{Basic theory}\label{sec:basics}

\subsection{Hamiltonian}\label{sec:Hamiltonian}

Zinc-blende 2DEGs grown along a crystal direction with at least two Miller indices equal in modulus enable the formation of persistent homogeneous and helical spin textures, the latter known as PSH.\cite{Kammermeier2016PRL}
Without loss of generality, we focus on a generic growth direction given by the unit vector $\hat{\bf n}$ lying in the first quadrant of the [110]-[001] crystal plane, i.e., $\hat{\bf n}=(\eta,\eta,\sqrt{1-2\eta^2})$ with $\eta\in[0,1/\sqrt{2}]$ and basis vectors  pointing along the high-symmetry crystal directions [100], [010], and [001].
For convenience, we define a coordinate system such that $\hat{\mb{x}}$ and $\hat{\mb{y}}$ axes span the conduction plane of the 2DEG while the $\hat{\mb{z}}$ axis corresponds to the quantum-well growth direction, i.e., $\hat{\mb{x}}=(n_z,n_z,-2\eta)/\sqrt{2}$, $\hat{\mb{y}}=(-1,1,0)/\sqrt{2}$, and $\hat{\mb{z}}\equiv\hat{\bf n}$.

In the vicinity of the $\Gamma$-point, the bandstructure of the 2DEG is described by the Hamiltonian
\begin{equation}
\mathcal{H}=\cfrac{\hbar^2k^2}{2m}+\cfrac{\hbar}{2}\,(\boldsymbol{\Omega}_1+\boldsymbol{\Omega}_3)\cdot\boldsymbol{\sigma}
\label{eq:Hamiltonian}
\end{equation}
with effective electron mass $m$, in-plane wave vector $\mb{k}=(k_x,k_y)$,
and the vector of Pauli matrices $\boldsymbol{\sigma}=(\sigma_x,\sigma_y,\sigma_z)$.
The SO coupling is characterized by the SO fields
\begin{align}
{\bf\Omega}_1&=\cfrac{2k}{\hbar}
\begin{pmatrix}
\left[\alpha+\beta^{(1)}(1+3\eta^2)n_z\right]\sin\varphi\\
\left[-\alpha+\beta^{(1)}(1-9\eta^2)n_z\right]\cos\varphi\\
-\sqrt{2}\beta^{(1)}\eta(1-3\eta^2)\sin\varphi
\end{pmatrix}
,\label{eq:SOF1}
\end{align}
and
\begin{align}
{\bf\Omega}_3&=\cfrac{2k}{\hbar}\beta^{(3)}
\begin{pmatrix}
(1-3\eta^2)n_z\sin3\varphi\\
-(1-3\eta^2)n_z\cos3\varphi\\
3\sqrt{2}\eta(1-\eta^2)\sin3\varphi
\end{pmatrix},
\label{eq:SOF3}
\end{align}
sorted in terms of first and third angular harmonics in the in-plane wave vector, represented in polar coordinates, i.e., $k_x=k\cos\varphi$ and $k_y=k\sin\varphi$, with in-plane polar angle $\varphi$.\cite{Kammermeier2016PRL}

The first angular harmonic SO field $\boldsymbol{\Omega}_1$ depends on the coefficients of Rashba and effective $k$-linear  Dresselhaus SO coupling, $\alpha=\gamma_{\rm R}\mathcal{E}_z$ and $\beta^{(1)}=\gamma_{\rm D}(\langle k_z^2\rangle - k^2/4)$, respectively. 
Both coefficients involve the material-specific bulk parameters $\gamma_{\rm R, D}$.
Assuming an approximately constant potential gradient along the quantum-well growth direction $\hat{\mb{z}}$, the Rashba coefficient scales linearly with the electric-field component $\mathcal{E}_z$. 
The effective $k$-linear Dresselhaus SO coefficient is predominantly determined by the width and structure of the quantum well through the projection $\langle k_z^2\rangle$ on the lowest bound state, which yields, for instance, $\langle k_z^2\rangle=(\pi/a)^2$ in an infinite square-well potential of width $a$.
Additionally, the coefficient $\beta^{(1)}$ includes a small term $\propto k^2$, which originates from the first angular harmonic part of the $k$-cubic Dresselhaus SO coupling. 
The third angular harmonic part of the $k$-cubic Dresselhaus SO coupling, yielding the field $\boldsymbol{\Omega}_3$, is distinguished by the prefactor $\beta^{(3)}=\gamma_{\rm D}k^2/4$.
As consequence of the proportionality $\propto k^2$, both Dresselhaus coefficients $\beta^{(1)}$ and $\beta^{(3)}$ are functions of the electron sheet density $n_s$, which at zero temperature is related to  the Fermi wave vector as $k_{\rm F}=\sqrt{2\pi n_s}$.
Since the first angular harmonic contribution of the $k$-cubic Dresselhaus SO coupling only renormalizes the magnitude of the $k$-linear Dresselhaus SO coupling, we follow common practice and denote $\boldsymbol{\Omega}_1$ as $k$-linear and  $\boldsymbol{\Omega}_3$ as $k$-cubic SO fields for simplicity.
In respect of D'yakonov-Perel' spin relaxation, it is practical to work with the ratios of the SO coefficients, for which we introduce the definitions $\Gamma_1=\alpha/\beta^{(1)}$ and $\Gamma_3=\beta^{(3)}/\beta^{(1)}$.

\subsection{Long-lived spin textures}\label{sec:SDE}

The spatiotemporal evolution of a spin density in presence of SO coupling has been intensively studied in different parameter regimes using semiclassical~\cite{Malshukov2000,Schwab2006,Yang2010,Luffe2011} or diagrammatic~\cite{Burkov2004,Stanescu2007,Wenk2010,Liu2012}  techniques.

We focus here on the diffusive regime with weak disorder and SO coupling, where the elastic mean free path is much longer than the Fermi wavelength but much shorter than spin-precession and all phase-breaking lengths (e.g. inelastic scattering, magnetic dephasing).
The scattering potentials are considered spin-independent and temperature effects only contribute to inelastic scattering processes by electron-phonon interaction.
Under these preconditions, the dynamic of a  local spin-density $\mb{s}(\mb{r},t)$ at position $\mb{r}$ and time $t$ can be computed by the spin-diffusion equation, which reads in reciprocal space\cite{wenkbook,Kammermeier2016PRL,Iizasa2020} 
\begin{align}
\frac{\partial }{\partial t}\tilde{\mb{s}}(\mb{q},t)=-\sdo(\mb{q})\;\tilde{\mb{s}}(\mb{q},t)
\label{eq:sde}
\end{align}
with wave vector $\mb{q}$ and Fourier-transformed spin-density $\tilde{\mb{s}}(\mb{q},t)=\int d^2r\; e^{-i\mb{q}\cdot\mb{r}}\mb{s}(\mb{r},t)$.
The spin-diffusion operator $\sdo$ for a generic 2DEG with the underlying Hamiltonian~(\ref{eq:Hamiltonian}) was computed in Refs.~\onlinecite{Kammermeier2016PRL,Iizasa2020} and its explicit form is 
presented in Appendix~\ref{app:sdo}.

Long-lived spin textures are determined by the minima in the spectrum $\lambda_l$ $(l\in\{0,\pm 1\})$ of the spin-diffusion operator $\sdo$, corresponding to the D'yakonov-Perel' spin-relaxation rates.
A special situation arises if the linear SO coefficients fulfill the relation
\begin{align}
\Gamma_1=\Gamma_0:=(1-9\eta^2)n_z
\label{eq:PSH_condition}
\end{align}
and the cubic Dresselhaus SO field vanishes, i.e., ${\Gamma_3=0}$.\cite{Kammermeier2016PRL}
(Note that the [001] growth direction has the additional solution with $\Gamma_1=-1$.)
Such configurations establish a PSH symmetry, in which the SO field ${\bf\Omega}_1$  (${\bf\Omega}_3=\mb{0}$) becomes collinear with the vector 
\begin{equation}
\hat{\bm{u}}_{\rm PSH}=\cfrac{{\rm sgn}(1-3\eta^2)}{\sqrt{2-3\eta^2}}
\begin{pmatrix}
-\sqrt{2}n_z\\
0\\
\eta
\end{pmatrix},
\label{eq:PSHfield}
\end{equation}
enabling the emergence of both homogeneous (homo) and helical (PSH) spin textures with infinite lifetime, that are,
\begin{align}
{\mb{s}}_{\rm homo}={}& \pm\hat{\mb{u}}_{\rm PSH},\label{eq:homo}\\
{\mb{s}}_{\rm PSH}={}&(\hat{\mb{y}}\times\hat{\mb{u}}_{\rm PSH})\cos({\mb{Q}}\cdot{\mb{r}}) - \hat{\mb{y}}\sin({\mb{Q}}\cdot{\mb{r}}), \label{eq:psh}
\end{align}
with spin-helix wave vector
\begin{equation}
\mb{Q}=Q_0\sqrt{1-3\eta^2/2}|1-3\eta^2|\;\hat{\mb{y}}\label{eq:Qpsh}
\end{equation}
and $Q_0 = 4m\beta^{(1)}/\hbar^2$.\cite{Kammermeier2016PRL,Iizasa2020}
The homogeneous spin texture is spin-polarized along the direction of the collinear SO field while the helical texture performs a rotation in real space perpendicular to the SO field axis.

For small parametric deviations from the ideal precondition, for which the SO-field collinearity is only weakly broken, the long-lived  spin textures retain the real-space structure (\ref{eq:homo})  and (\ref{eq:psh}) but decay exponentially with the characteristic lifetimes $\tau_\text{homo}$ and $\tau_\text{PSH}$.\cite{Kammermeier2016PRL,Iizasa2020}
Here, the spectrum of $\sdo$ can be approximated as
\begin{align}
\lambda_l={}&D_e\left[q_x^2+(q_y+l\, Q)^2\right]+\tilde{\Delta}_l, \quad l\in\{0,\pm 1\},
\label{eq:SDspectrumPSH}
\end{align}
where $\tilde{\Delta}_{\pm 1}=1/\tau_{\rm PSH}$ and $\tilde{\Delta}_{0}=1/\tau_{\rm homo}$ denote the spin-relaxation rates
\begin{align}
\cfrac{1}{\tau_{\rm PSH}}&=\cfrac{4\pi^2}{\tau_0}\bigg[\cfrac{3 - 17 \eta^2 + 85 \eta^4 - 171 \eta^6 + 108 \eta^8}{8 - 12 \eta^2}\Gamma_3^2\notag\\
&\phantom{=\cfrac{4\pi^2}{\tau_0}\bigg[}+\frac{1}{8}(\Gamma_1-\Gamma_0)^2\bigg],\label{eq:PSHrate}\\
\frac{1}{\tau_{\rm homo}}&=\cfrac{4\pi^2}{\tau_0}\bigg[\cfrac{n_z^2(1+17 \eta^2 -45 \eta^4 +27 \eta^6 )}{4 - 6 \eta^2}\Gamma_3^2\notag\\
&\phantom{=\cfrac{4\pi^2}{\tau_0}\bigg[}+\frac{1}{4}(\Gamma_1-\Gamma_0)^2\bigg].
\label{eq:HOMOrate}
\end{align}
In expressions (\ref{eq:PSHrate}) and (\ref{eq:HOMOrate}) we combined the results of Refs.~\onlinecite{Kammermeier2016PRL,Iizasa2020} under the assumption that the deviations from PSH symmetry induced by $k$-linear and $k$-cubic SO terms are small and independent of each other, i.e., $\Gamma_1$ being close to $\Gamma_0$ and $\Gamma_3\ll 1$.
Additionally, it is required that the magnitude of the SO field ${\bf\Omega}_3$ should much smaller than ${\bf\Omega}_1$.
Even though this is normally the case, for growth directions very near [111] the Rashba and $k$-linear Dresselhaus SO couplings suppress each other and the total SO field can strongly differ from collinearity.

The particular form of the spin-diffusion spectrum (\ref{eq:SDspectrumPSH}) allows us  to derive a closed-form expression for the magneto-conductivity correction due to weak antilocalization in Sec.~\ref{sec:SOCaab}.

\subsection{Weak (Anti)localization}\label{sec:WAL}

The weak (anti)localization correction to the longitudinal conductivity in the diffusive regime, analogous to the spin-diffusion equation in Sec.~\ref{sec:SDE}, is given by\cite{Hikami1980}
\begin{align}
\Delta\sigma={}&-\frac{2e^2}{h}\frac{\hbar D_e}{\mc{V}}\sum_{\gamma,\delta}\sum_\mb{q}\braket{\gamma\delta|\;\mc{C}(\mb{q})\;|\delta\gamma},
\label{eq:WAL1}
\end{align} 
where $D_e$ is the 2D diffusion constant, $e>0$ the elementary charge, $\mc{V}$ the 2D volume, $\gamma,\delta\in\{\pm 1/2\}$ the spin-1/2 indices and $\mb{q}\perp \mbs{\hat{z}}$ the total in-plane wave vector of the two interfering electron waves.
The Cooperon $\mc{C}$ (in units of 1/energy) describes the propagation of two electrons traveling along closed-loop trajectories in opposite directions.\cite{AkkermansBook}
The derivation of this expression is reviewed in detail, e.g., in Ref.~\onlinecite{KammermeierPHD}.

If Zeeman spin splitting is negligible, it is convenient to select the singlet-triplet representation $\ket{j,m_j}$ for total spin-1 particles, with total-spin quantum number $j\in\{0,1\}$ and magnetic quantum number $m_j\in\{0,\pm 1\}$, as the singlet $(j=0)$ and triplet $(j=1)$ sectors decouple.
In this representation, we can express $\Delta\sigma$ in terms of the  singlet  and triplet  eigenvalues, $E^S$ and $E^T_l$ ($l\in\{0,\pm 1\}$) of the inverse Cooperon $\mc{C}^{-1}$, which reads
\begin{align}
\Delta\sigma={}&\frac{2e^2}{h}\frac{\hbar D_e}{\mc{V}}\sum_\mb{q}\left[\frac{1}{E^S(\mb{q})}-\sum_{l}\frac{1}{E^T_l(\mb{q})}\right].
\label{eq:WAL2}
\end{align} 
While the singlet eigenvalue can be generally  written as $E^S=\hbar D_e q^2$, the triplet eigenvalues $E^T_l$ are sensitive to the SO coupling as it mixes the triplet components.
Without magnetic field, the triplet sector of the inverse Cooperon $\mc{C}^{-1}$ is linked to the spin-diffusion operator $\sdo$ by a unitary transformation $U$ via 
\begin{align}
\mc{C}^{-1}= \hbar\,U \sdo U^\dag,
\label{eq:Cooperon_relation}
\end{align}
where $U$ relates the triplet basis to spin-density components (cf. Appendix~\ref{app:SDO_Cooperon} for more details).\cite{Wenk2010}
Therefore, the triplet eigenvalues $E^T_l$ follow directly from the diagonalization of $\sdo$, which is given for a generic 2DEG with at least two growth-direction Miller indices equal in modulus in Eq.~(\ref{eq:sdo}).

In 2D, the sum over $\mb{q}$ diverges as $\ln (q)$ for small and large wave vectors.
The standard regularization procedure consists of selecting cut-off parameters for small and large $q$.
The upper limit $q_\text{max}$ is naturally given by the shortest diffusion, which means a single collision, where the distance is given by the mean free path $l$.
It is common to select $q_\text{max}=1/\sqrt{D_e\tau}$, where $\tau$ is the elastic scattering rate.
\footnote{Although $\sqrt{D_e\tau}=l/\sqrt{2}$ is smaller than the mean free path $l$, in the diffusive regime the exact value of the upper cut-off hardly affects the final result.}
The lower limit is determined by the finite dephasing times due to inelastic scattering ($\tau_\phi$) and magnetic field ($\tau_B$).
This cut-off is usually introduced as additional positive energy shifts $\hbar/\tau_{\phi,B}$ in the eigenvalues $E^S$ and $E^T_l$ of the inverse Cooperon.
In the sum over $\mb{q}$, the vanishing wave vectors can then be included without causing divergence.

\subsection{Magneto-conductivity}\label{sec:MC}

Experimentally, the weak (anti)localization correction is typically probed in response of a weak out-of-plane magnetic field $\mb{B}\parallel\hat{\mb{z}}$, where Zeeman spin splitting is negligible.
The orbital contribution of the magnetic field suppresses the interference of electrons counter-propagating along time-reversed paths by breaking of time-reversal invariance and introducing an Aharonov-Bohm phase.
It is, therefore, practical to define the \textit{relative} magneto-conductivity  correction  
\begin{align}
\Delta\sigma_R(B):=&\Delta\sigma(B)-\Delta\sigma(0),
\end{align}
with magnetic-field strength $B:=\Vert\mb{B}\Vert$, that constitutes the experimentally measured quantity.

In this paper, we juxtapose two distinct methods for the inclusion of the phase-breaking effect of a weak perpendicular magnetic field.

\subsubsection{Landau-level picture}

In the Landau-level (ll) picture, the magnetic field restricts the electron motion to highly energy-degenerate Landau orbitals.
Accordingly, the in-plane wave-vector components $q_{x,y}$ are expressed in terms of ladder operators $a^\dag$ and $a$ as 
\begin{align}
q_x=(a+a^\dag)/\sqrt{2}l_B,\quad
q_y=(a-a^\dag)/i\sqrt{2}l_B
\label{eq:ladder_op}
\end{align}
with the magnetic length $l_B=\sqrt{\hbar/2eB}$ of a particle with effective charge $2e$.
The ladder operators $a^\dag(a)$ raise (lower) the index of a Landau level  $\ket{n}$ with $ n\in \mathbb{N}_0$ and thereby lead to the non-vanishing matrix elements
\begin{align}
\braket{n-1|a|n}=\braket{n|a^\dag|n-1}=\sqrt{n},\quad\braket{n|a^\dag a|n}=n
\label{eq:ladder_op_matrix_el}
\end{align}
with $[a,a^\dag]=1$ and $a\ket{0}=0$.
Thus, the sum over in-plane wave-vector $\mb{q}$ in Eqs.~(\ref{eq:WAL1}) and (\ref{eq:WAL2}) is replaced by a trace over Landau levels $n$ multiplied by the degeneracy factor $g=\mc{V}/2\pi l_B^2$, i.e.,
\begin{align}
\Delta\sigma^{(ll)}={}&\frac{e^2}{2\pi^2\hbar}\frac{\hbar D_e}{ l_B^2}\sum_{n=0}^{N_\text{max}}\braket{n|\left[\frac{1}{E^S(\mb{q})}-\sum_{l}\frac{1}{E^T_l(\mb{q})}\right]|n}.
\label{eq:WAL3}
\end{align} 
The positive energy shift of $\mc{C}^{-1}$  in the lowest Landau level $\ket{0}$ without SO coupling can be interpreted as a magnetic-dephasing cut-off  
\begin{align}
\frac{\hbar}{\tau_B}:=\braket{0|\mc{C}^{-1}|0}=\braket{0|\hbar D_e q^2|0}=2eD_eB
\label{eq:tauB}
\end{align}
entering each singlet and triplet channel. 
Moreover, using the relation $\braket{n|q^2|n}=(2n+1)/l_B^2\leq 1/D_e\tau$, we identify the upper cut-off $N_\text{max}$ in the sum over Landau levels as
\begin{align}
N_\text{max}\approx\hbar/4eBD_e\tau=\tau_B/2\tau,
\label{eq:Nmax}
\end{align}
where the integer part has to be taken and we assumed $N_\text{max}\gg 1\Leftrightarrow \tau_B\gg\tau$ according to the diffusive limit.
This cut-off is equivalent to the one used in Refs.~\onlinecite{Pikus1995,Knap1996}.
Although the maximum Landau level has to be an integer, we can treat $N_\text{max}$ formally as a continuous parameter whenever an analytic result for the finite sum over Landau levels exists.

The $k$-linear SO field ${\bf\Omega}_1$ yields contributions in the Cooperon triplet sector that are linear in the wave-vector components $q_{x,y}$  (cf. the related spin-diffusion operator $\sdo$ in Eq.~\ref{eq:sdo}).
It follows from the relations (\ref{eq:ladder_op}) and (\ref{eq:ladder_op_matrix_el}) that the $q$-linear terms induce a mixing of Landau levels, which usually inhibits the derivation of an explicit solution for the Cooperon eigenvalues.
Thus, the diagonalization of the $4(N_\text{max}+1)$-dimensional Cooperon,
represented in the joint Landau-spin-1 basis $\ket{n}\ket{j,m_j}$, constitutes the key obstacle for finding an analytic solution for the  the magneto-conductivity correction.

\subsubsection{Quasi-classical approximation}\label{sec:quasi-class}

The quasi-classical (qc) approximation assumes that the magnetic field solely induces a dephasing rate $1/\tau_B= D_e/l_B^2$ given by the Aharonov-Bohm phase difference of two electrons counter-propagating in a closed loop.\cite{Beenakker1991}
The dephasing rate yields a positive energy shift $\hbar/\tau_B$ of the inverse Cooperon equivalent to the eigenvalue $E^S$ in the lowest Landau orbital $\ket{0}$ [cf. Eq.~(\ref{eq:tauB})].
Since effects of Landau quantization in the Cooperon spectrum are neglected otherwise, the wave-vector components $q_{x,y}$ remain quasi-continuous variables.
Thus, the sum over $\mb{q}$ in $\Delta\sigma$, Eq.~(\ref{eq:WAL2}), can be replaced by a 2D integral in $\mb{q}$-space, i.e., 
\begin{align}
\Delta\sigma^{(qc)}={}&\frac{e^2}{2\pi^2\hbar}\int_0^{2\pi} \frac{d\phi}{2\pi} \int_0^{1/\sqrt{D_e\tau}} dq\;q\left[\frac{\hbar D_e}{E^S(\mb{q})+\hbar/\tau_B}\right.\notag\\
&\phantom{}\left.-\sum_{l}\frac{\hbar D_e}{E^T_l(\mb{q})+\hbar/\tau_B}\right],
\label{eq:WAL4}
\end{align} 
where we expressed $\mb{q}$ in polar coordinates with polar angle $\phi$ and introduced an upper cut-off in the wave-vector integral due to elastic scattering.

\section{Spin-degenerate case}\label{sec:noSOC} 
 
To elucidate the differences in outcomes of Landau-quantization and quasi-classical approaches,
let us first study the simple situation of vanishing SO coupling.
Here, the singlet and triplet eigenvalues of $\mc{C}^{-1}$ are identical, i.e., $E^S=E^T_l=\hbar D_eq^2+\hbar/\tau_\phi$, where we included  an energy shift due to inelastic scattering.

\subsection{Quasi-classical approximation}
 
According to Eq.~(\ref{eq:WAL4}) the magneto-conductivity correction takes the form
\begin{align}
\Delta\sigma^{(qc)}{}&=-\frac{e^2}{\pi^2\hbar}\int_0^{2\pi} \frac{d\phi}{2\pi} \int_0^{1/\sqrt{D_e\tau}} dq \;\frac{D_e q}{D_eq^2+\tau_\phi^{-1}+\tau_B^{-1}}\notag\\
&=-\frac{e^2}{2\pi^2\hbar}\nu(0)\label{eq:WALnoSOC_qc},
\end{align}
with the function 
\begin{align}
\nu(x)={}&\ln\left(\frac{\tau^{-1}+\tau_\phi'(x)^{-1}}{\tau_\phi'(x)^{-1}}\right),\label{eq:nu}
\end{align}
depending on the \textit{total} dephasing rate
\begin{align}
\frac{1}{\tau_\phi'}(x) ={}&\frac{1}{\tau_\phi}+\frac{1}{\tau_B}+x.
\label{eq:tot_deph_rate}
\end{align}
In the diffusive limit, where $\tau\ll x^{-1},\tau_{\phi,B}$, we can simplify the function (\ref{eq:nu}) as
\begin{align}
\nu(x)\approx{}&\ln\left(\frac{\tau_\phi'(x)}{\tau}\right).
\label{eq:nu_approx}
\end{align} 
For $B=0$, Eq.~(\ref{eq:WALnoSOC_qc}) agrees with the well-known expression for the zero-field weak localization.~\cite{Iordanskii1994,Pikus1995}

Notice that the regularization of the logarithm in Eq.~(\ref{eq:nu}) is physically meaningful: the lower limit constitutes the \textit{total} dephasing rate, the upper limit the \textit{total} scattering rate, which according to Matthiessen's rule is the sum of all contribution scattering rates.
If we alternatively included the total dephasing as lower limit of the integral as in Ref.~\onlinecite{Marinescu2019}, the upper cut-off is solely given by the elastic scattering rate.
In the limit $\tau_{\phi,B}\gg\tau$, both approaches are equivalent, though.

\begin{figure}[t]
\includegraphics[scale=.5]{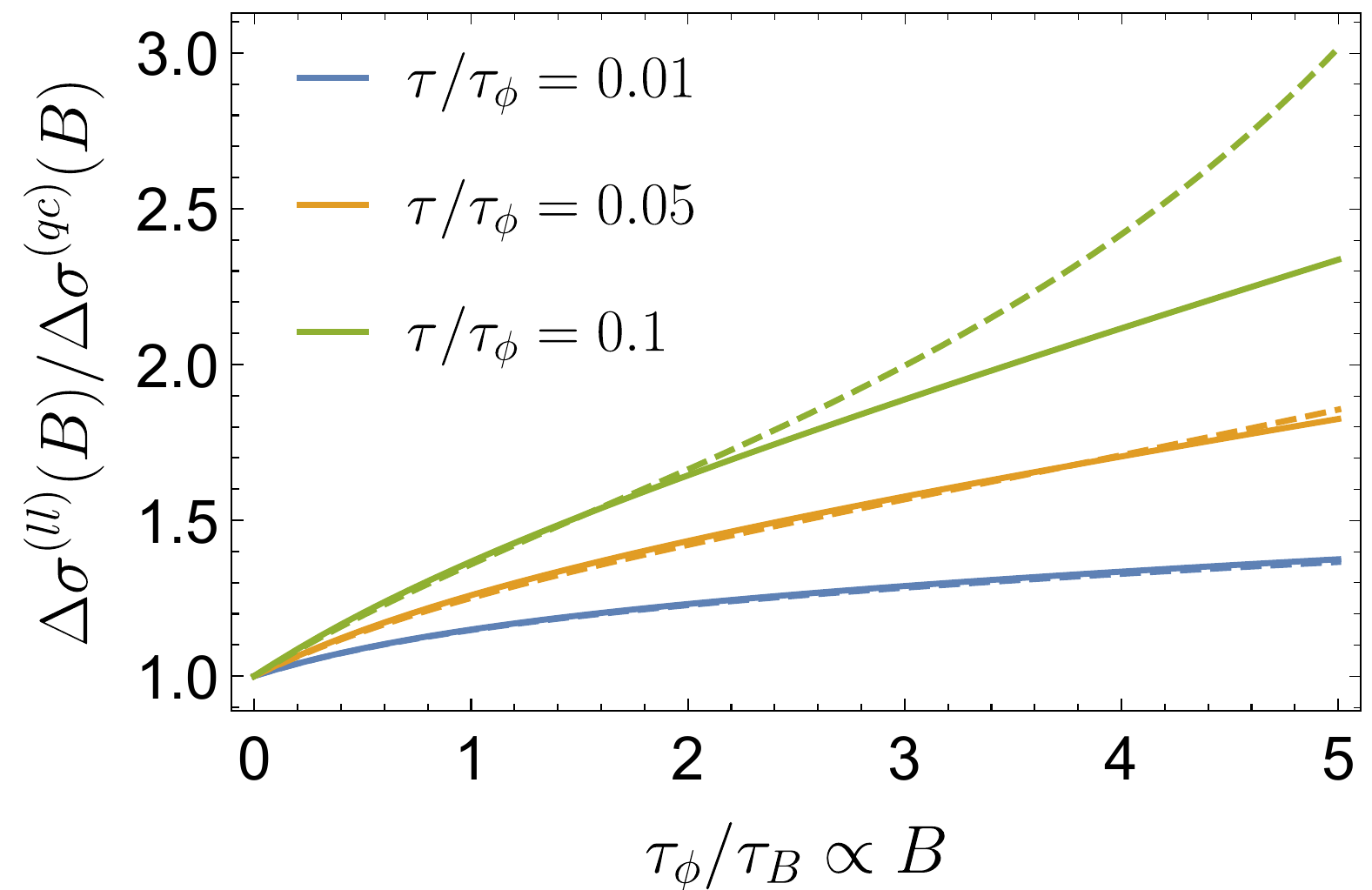}
\caption{Ratio of magneto-conductivity correction in Landau-level and quasi-classical picture $\Delta\sigma^{(ll)}/\Delta\sigma^{(qc)}$ in dependence of the ratio of inelastic scattering and magnetic dephasing times $\tau_\phi/\tau_B$, proportional to the magnetic-field strength $B$, without SO coupling.
The solid lines use the expressions (\ref{eq:WALnoSOC_qc}) and (\ref{eq:WALnoSOC_ll}), while the dashed lines use approximations Eq.~(\ref{eq:nu_approx}) and (\ref{eq:xi_approx}),  valid in the diffusive limit, where $\tau\ll\tau_{\phi,B}$.
}
\label{fig:ratio_sigma_qc_ll}
\end{figure}

\subsection{Landau-level picture}

Taking into account the Landau quantization, $\Delta\sigma$ is given by  Eq.~(\ref{eq:WAL3}) and reads
\begin{align}
\Delta\sigma^{(ll)}
&={}-\frac{e^2}{2\pi^2\hbar}\sum_{n=0}^{N_\text{max}}\left(n+\frac{1}{2}+\frac{\tau_B}{2\tau_\phi}\right)^{-1}\notag\\
&={}-\frac{e^2}{2\pi^2\hbar}\xi(0)\label{eq:WALnoSOC_ll},
\end{align} 
where 
\begin{align}
\xi(x)={}&\Psi\left(\frac{3}{2}+\frac{\tau_B}{2}\bigg[\frac{1}{\tau}+\frac{1}{\tau_\phi}+x\bigg]\right)\notag\\
&-\Psi\left(\frac{1}{2}+\frac{\tau_B}{2}\bigg[\frac{1}{\tau_\phi}+x\bigg]\right)\label{eq:xi}
\end{align}
with the Digamma function $\Psi$, being the logarithmic derivative of the Gamma function.
Using the approximation 
\begin{align}
\xi(x)\approx{}&\ln\left(\frac{\tau_B}{2\tau}\right)-\Psi\left(\frac{1}{2}+\frac{\tau_B}{2}\bigg[\frac{1}{\tau_\phi}+x\bigg]\right),\label{eq:xi_approx}
\end{align} 
in the diffusive limit, $\tau\ll x^{-1},\tau_{\phi,B}$, where we employed the asymptotic expansion of the Digamma function $\Psi(x)\approx \ln{x}$ for $x\rightarrow\infty$, Eq.~(\ref{eq:WALnoSOC_ll}) agrees with the result in Ref.~\onlinecite{Bergmann1984}.
Notice that in other papers the expressions are often transformed in magnetic-field scales, e.g., $\tau_B/2\tau_\phi=B_\phi/B$ where $B_\phi=\hbar/4eD_e\tau_\phi$.

Without magnetic field, where $\tau_B\rightarrow \infty$, the function $\xi$ becomes identical to $\nu$ and $\sigma^{(qc)}$ equals $\sigma^{(ll)}$.
In the following, we see, however, that for finite magnetic fields differences arise even in lowest order in $B$.

\subsection{Comparison}

Initially, we turn to the numerical comparison of the results obtained by the quasi-classical and Landau-level approaches.
According to the diffusive regime, the elastic scattering time should be much shorter than the phase-breaking time scales, i.e., $\tau\ll\tau_{\phi,B}$.
Moreover, the effect of the magnetic field becomes appreciable if $\tau_B<\tau_\phi$.

\begin{figure}[t]
\includegraphics[scale=.5]{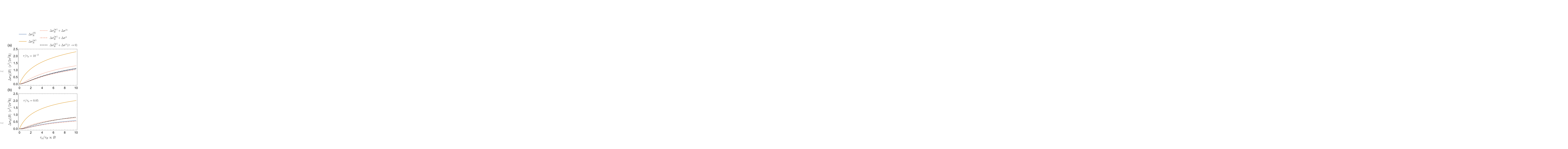}
\caption{Relative magneto-conductivity correction in the Landau-level (blue) and quasi-classical picture (yellow)  for (a) $\tau/\tau_\phi=10^{-2}$ and (b) $\tau/\tau_\phi=0.05$ in dependence of the ratio of inelastic scattering and magnetic dephasing times $\tau_\phi/\tau_B$ (proportional to the magnetic-field strength $B$) without SO coupling.
The solid lines are obtained using the expressions (\ref{eq:WALnoSOC_qc}) and (\ref{eq:WALnoSOC_ll}).
The red dotted line uses the quasi-classical approximation with inclusion of the first Euler-MacLaurin correction $\Delta\sigma^{\chi_1}$ [Eq.~(\ref{eq:chi12})], the red dot-dashed line the first two corrections $\Delta\sigma^{\chi}$ [Eqs.~(\ref{eq:chi})], and the black dashed line the first two corrections but considers the diffusive limit, where $\tau\ll\tau_{\phi,B}$.
}
\label{fig:sigma_R_noSOC}
\end{figure}

In Fig.~\ref{fig:ratio_sigma_qc_ll}, we plot the ratio of magneto-conductivity correction in the Landau-level and quasi-classical picture $\Delta\sigma^{(ll)}/\Delta\sigma^{(qc)}$ in dependence of the ratio of inelastic scattering and magnetic dephasing times $\tau_\phi/\tau_B$, which is proportional to the magnetic-field strength $B$.
Depending on $B$ and the ratio $\tau/\tau_\phi$, the deviation from unity can be significant.
This becomes particularly apparent when comparing the corresponding \textit{relative} magneto-conductivity corrections $\Delta\sigma_R$ (blue and yellow lines) in Fig.~\ref{fig:sigma_R_noSOC}.
Interestingly, while in the moderate magnetic-field regime, where $\tau_\phi\gg\tau_B$ both functions exhibit qualitatively the same trend, in the limit $B\rightarrow 0$ the tangents grossly differ, leading to a large separation of both curves for small magnetic fields.

To better understand the differences, we inspect the asymptotic limits of small and moderate magnetic fields.

\subsubsection{Small magnetic fields: $\tau_B\gg\tau_\phi\gg\tau$}

We find to second order in the magnetic field (${1/\tau_B\propto B}$)
\begin{align}
\Delta\sigma^{(qc)}&\rightarrow{}-\frac{e^2}{2\pi^2\hbar}\left[\ln \left(\frac{\tau_\phi}{\tau}\right)-\frac{\tau_\phi}{\tau_B}+\frac{\tau_\phi^2}{2\tau_B^2}\right]
\label{eq:WALnoSOC_qc_assympt_1},\\
\Delta\sigma^{(ll)}&\rightarrow{}-\frac{e^2}{2\pi^2\hbar}\left[\ln \left(\frac{\tau_\phi}{\tau}\right)-\frac{\tau_\phi^2}{6\tau_B^2}\right]
\label{eq:WALnoSOC_ll_assympt_1},
\end{align} 
which gives to \textit{lowest} order in the magnetic field
\begin{align}
\Delta\sigma_R^{(qc)}&\rightarrow{}\frac{e^2}{2\pi^2\hbar}\;\frac{\tau_\phi}{\tau_B}
\label{eq:WALnoSOC_qc_rel_assympt_1},\\
\Delta\sigma_R^{(ll)}&\rightarrow{}\frac{e^2}{2\pi^2\hbar}\;\frac{\tau_\phi^2}{6\tau_B^2}
\label{eq:WALnoSOC_ll_rel_assympt_1}.
\end{align} 
Since for small magnetic fields the logarithmic function in $\Delta\sigma$ dominates, the ratio $\Delta\sigma^{(ll)}/\Delta\sigma^{(qc)}$ in Fig.~\ref{fig:ratio_sigma_qc_ll} approaches unity for decreasing $\tau/\tau_\phi$ and $B$.
Yet, the magneto-conductivity exhibits a different scaling with the magnetic field in lowest order (linear and quadratic), which explains the $B$-linear increase for $B\rightarrow 0$ in Fig.~\ref{fig:ratio_sigma_qc_ll} and the large discrepancy in Fig.~\ref{fig:sigma_R_noSOC}.
The asymptotic functions for $\Delta\sigma_R$ are displayed as dotted lines in Fig.~\ref{fig:sigma_R_noSOC_asympt}(a), which corresponds to Fig.~\ref{fig:sigma_R_noSOC}(a) with focus on low values of $\tau_\phi/\tau_B$.
The low-field behavior of $\Delta\sigma_R^{(ll)}$ was already mentioned by Hikami \textit{et al.}\cite{Hikami1980}
Noteworthy, although the underlying mechanisms may differ, other authors also reported a $B$-parabolic scaling in the low-field  magneto-conductivity in thin films, wires, and quantum dots.\cite{Dugaev1984,Meyer2002,Brouwer2002,Cremers2003}

\subsubsection{Moderate magnetic fields: $\tau_\phi\gg\tau_B\gg\tau$}

\begin{figure}[t]
\includegraphics[scale=.5]{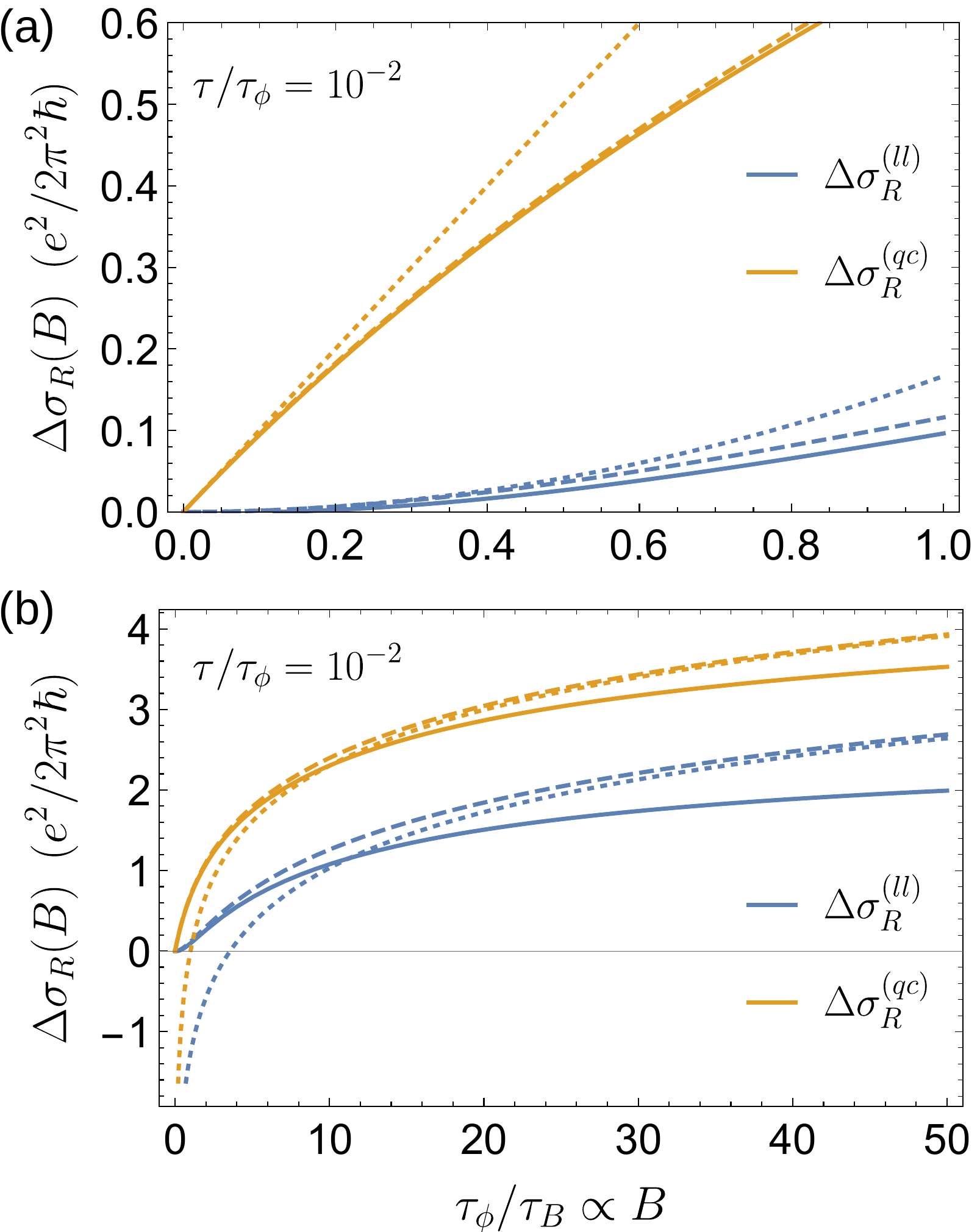}
\caption{Relative magneto-conductivity correction in the Landau-level (blue) and quasi-classical picture (yellow)  for the asymptotic limits of (a) small magnetic fields ($\tau_B\gg\tau_\phi\gg\tau$) and (b) moderate magnetic fields ($\tau_\phi\gg\tau_B\gg\tau$) in dependence of the ratio of inelastic scattering and magnetic dephasing times $\tau_\phi/\tau_B$ (proportional to the magnetic-field strength $B$) without SO coupling.
The solid lines are obtained using the exact expressions (\ref{eq:WALnoSOC_qc}) and (\ref{eq:WALnoSOC_ll}), while the dashed lines use the simplified form using the approximations Eq.~(\ref{eq:nu_approx}) and (\ref{eq:xi_approx}),  valid in the diffusive limit, where $\tau\ll\tau_{\phi,B}$.
The dotted lines correspond to the asymptotic expressions (\ref{eq:WALnoSOC_qc_rel_assympt_1})--(\ref{eq:WALnoSOC_ll_rel_assympt_2}). }
\label{fig:sigma_R_noSOC_asympt}
\end{figure}

For moderate magnetic fields, we obtain
\begin{align}
\Delta\sigma_R^{(qc)}&\rightarrow{}\frac{e^2}{2\pi^2\hbar}\ln \left(\frac{\tau_\phi}{\tau_B}\right)
\label{eq:WALnoSOC_qc_rel_assympt_2},\\
\Delta\sigma_R^{(ll)}&\rightarrow{}\frac{e^2}{2\pi^2\hbar}\left[\ln \left(\frac{\tau_\phi}{\tau_B}\right)+\ln2+\Psi(1/2)\right]
\label{eq:WALnoSOC_ll_rel_assympt_2}.
\end{align} 
While here the $B$-dependence is logarithmic in both quasi-classical approximation and Landau-level picture, a quantitative agreement is only found if $\ln (\tau_\phi/\tau_B)\gg\vert\ln2+\Psi(1/2)\vert\approx 1.3$.
This requirement is difficult to fulfill in experiment since  $\tau\ll\tau_B$ must hold at the same time. 
For instance, even for $\tau_\phi/\tau_B=10^{2}$ we obtain only  $\ln(\tau_\phi/\tau_B)\approx 4.6$, which is not considerably larger than 1.3.
The asymptotic functions for $\Delta\sigma_R$ are displayed as dotted lines in Fig.~\ref{fig:sigma_R_noSOC_asympt}(b), which corresponds to Fig.~\ref{fig:sigma_R_noSOC}(a) with focus on moderate values of $\tau_\phi/\tau_B$.

\subsection{Euler-MacLaurin corrections to the quasi-classical approximation}

Comparing the magneto-conductivity corrections in expressions  (\ref{eq:WALnoSOC_qc}) and (\ref{eq:WALnoSOC_ll}), it becomes clear that $\Delta\sigma^{(qc)}$ is obtained from $\Delta\sigma^{(ll)}$ by  performing the substitution $n\rightarrow  q^2 D_e\tau_B/2$ and replacing the sum over Landau levels by an integral over the wave-vector. 
Since integration and summation are mathematically not identical, we explore whether a more accurate agreement can be obtained using corrections given by the Euler-MacLaurin formula, which describes the differences between sum and integral.\cite{DevriesBook1994}
Details about this formula are given in Appendix \ref{app:sum_integral}.
We find that adding the leading two corrections $\Delta \sigma^{\chi_{1}}$ and $\Delta \sigma^{\chi_{2}}$ to the quasi-classical formula $\Delta\sigma^{(qc)}$ for the magneto-conductivity correction provides an excellent approximation of $\Delta\sigma^{(ll)}$ without performing the explicit sum over Landau levels.
The correction terms are%
\begin{align}
\Delta \sigma^{\chi_{1,2}}&={}-\frac{e^2}{2\pi^2\hbar}f_{1,2}(0)
\label{eq:chi12}
\end{align}
with the functions
\begin{align}
f_1(x)={}&\frac{\tau_\phi'(x)\big[2\tau+\tau_\phi'(x)\big]}{\tau_B\big[\tau+\tau_\phi'(x)\big]},\label{eq:f1}\\
f_2(x)={}&\frac{\tau_\phi'(x)^3\big[2\tau+\tau_\phi'(x)\big]}{3\tau_B^2\big[\tau+\tau_\phi'(x)\big]^2}\label{eq:f2}
\end{align}
and the total dephasing rate $1/\tau_\phi'(x)$ as defined in Eq.~(\ref{eq:tot_deph_rate}).
Adding both corrections, we define 
\begin{align}
\Delta \sigma^{\chi}:={}\Delta \sigma^{\chi_1}+\Delta \sigma^{\chi_2}={}-\frac{e^2}{2\pi^2\hbar}f(0),
\label{eq:chi}
\end{align}
with $f=f_1+f_2$, which can be approximated  as
\begin{align}
f(x)\approx{}&\frac{\tau_\phi'(x)}{\tau_B}+\frac{\tau_\phi'(x)^2}{3\tau_B^2}
\label{eq:f_approx}
\end{align} 
in the diffusive limit, where $\tau\ll x^{-1},\tau_{\phi,B}$.

For small magnetic fields ($\tau_B\gg\tau_\phi$), we thus have in the diffusive limit
\begin{align}
\Delta \sigma^{\chi}&\rightarrow{}-\frac{e^2}{2\pi^2\hbar}\left(\frac{\tau_\phi}{\tau_B}-\frac{2\tau_\phi^2}{3\tau_B^2}\right),
\end{align}
which when added to $\Delta \sigma^{(qc)}$ [Eq.~(\ref{eq:WALnoSOC_qc_assympt_1})] equals $\Delta \sigma^{(ll)}$ [Eq.~(\ref{eq:WALnoSOC_ll_assympt_1})] in the Landau-level picture.
More specifically, the first correction cancels the $B$-linear term, the
second correction allows to recover the accurate parabolicity.
Hence, including the first two corrections of the Euler-MacLaurin formula to the integral approximation in the quasi-classical approach enables to restore the appropriate low-field scaling of the magneto-conductivity given by the Landau-level picture.

With regard to the moderate-magnetic-field regime ($\tau_B\ll\tau_\phi$), we obtain  in the diffusive limit
\begin{align}
\Delta \sigma^{\chi}&\rightarrow{}-\frac{e^2}{2\pi^2\hbar}\;\frac{4}{3}.
\end{align}
Adding this term to $\Delta \sigma^{(qc)}$ [Eq.~(\ref{eq:WALnoSOC_qc_rel_assympt_2})] closes the conductivity shift compared to $\Delta \sigma^{(ll)}$ in Eq.~(\ref{eq:WALnoSOC_ll_rel_assympt_2}) since 
$\ln 2+\Psi(1/2)\approx -4/3$.

In Figure \ref{fig:sigma_R_noSOC}, we display the impact of the Euler-MacLaurin corrections to the relative magneto-conductivity correction  $\Delta \sigma_R^{(qc)}$ in quasi-classical approximation (yellow solid line)  for different ratios of $\tau/\tau_\phi$.
The red dotted lines include the first correction $\Delta \sigma^{\chi_1}$, the red dot-dashed lines the first two corrections $\Delta \sigma^{\chi}$, and the black dashed line the first two corrections $\Delta \sigma^{\chi}$ with the approximation of function $f$ in diffusive limit, Eq.~(\ref{eq:f_approx}).
While the first correction $\Delta \sigma^{\chi_1}$ provides already large enhancement of the magneto-conductivity profile, excellent agreement with the exact result $\Delta \sigma_R^{(ll)}$ (blue solid line) is found when both Euler-MacLaurin corrections are taken into account.
The accuracy of the approximation of $\Delta \sigma^{\chi}$ in the diffusive limit (black-dashed line) increases with reduction of the ratio $\tau/\tau_\phi$.

\section{(001)-2DEG with spin-splitting}\label{sec:SOC001}

In the following, we turn to the more intricate scenario of a zinc-blende 2DEG with both Rashba and Dresselhaus SO coupling and account for the arising D'yakonov-Perel' spin relaxation.~\cite{perel} 
For a quantum-well grown along the [001] high-symmetry crystal direction, the triplet sector of the inverse Cooperon can be written as
\begin{align}
\frac{\mc{C}^{-1}}{D_e\hbar}=&{}\;q^2+  q_x Q_0(1-\Gamma_1)S_y+  q_y  Q_0(1+\Gamma_1)S_x\notag\\
&+\frac{Q_0^2}{4}  \big\{\left[(1+\Gamma_1)^2+\Gamma_3^2\right]S_x^2 \notag\\
&\phantom{+\frac{Q_0^2}{4}\big[\; }+ \left[(1-\Gamma_1)^2+\Gamma_3^2\right]S_y^2\big\},
\label{eq:Cooperon001}
\end{align}
where we used its relation (\ref{eq:Cooperon_relation}) to the spin-diffusion operator $\sdo$, Eq.~(\ref{eq:sdo}), which is evaluated at $\eta=0$ and expanded in spin-1 basis matrices $S_{x,y,z}$, listed in Appendix~\ref{app:spin_matrices}.

As a result of the SO coupling, the Cooperon is generally non-diagonal in the spin-1 triplet subspace.
In addition, the Rashba and effective $k$-linear Dresselhaus SO  field $\mbs{\Omega}_1$, Eq.~(\ref{eq:SOF1}), yields contributions in the inverse Cooperon that are also linear in the wave vector $q_{x,y}$ and thus induce a mixing of adjacent Landau levels, i.e., in general we have $\braket{n \pm 1|\mc{C}|n}\neq 0$. 
The latter property makes an analytic treatment particularly challenging as the simple association $n=q^2 l_B^2/2$ between wave-vector and Landau-level quantum number is not possible for linear $q$.
Consequently, exact solutions exist only in a very limited number of scenarios: (i) pure $k$-cubic Dresselhaus SO coupling (${\alpha=\beta^{(1)}=0}$), (ii) either Rashba ($\beta^{(1,3)}=0$) or Dresselhaus SO coupling ($\alpha=0$), or (iii) for PSH symmetry ($\Gamma_1=\pm 1$ and $\Gamma_3=0$).

In the following, we briefly review all three cases separately.
The last case is then used in the next section
as basis to obtain an approximate closed-form expression for the magneto-conductivity correction of an arbitrary confined 2DEG near the PSH-symmetry point.

\subsection{Pure cubic Dresselhaus SO coupling}\label{sec:pureCubicDress}
\begin{figure}[t]
\includegraphics[scale=.5]{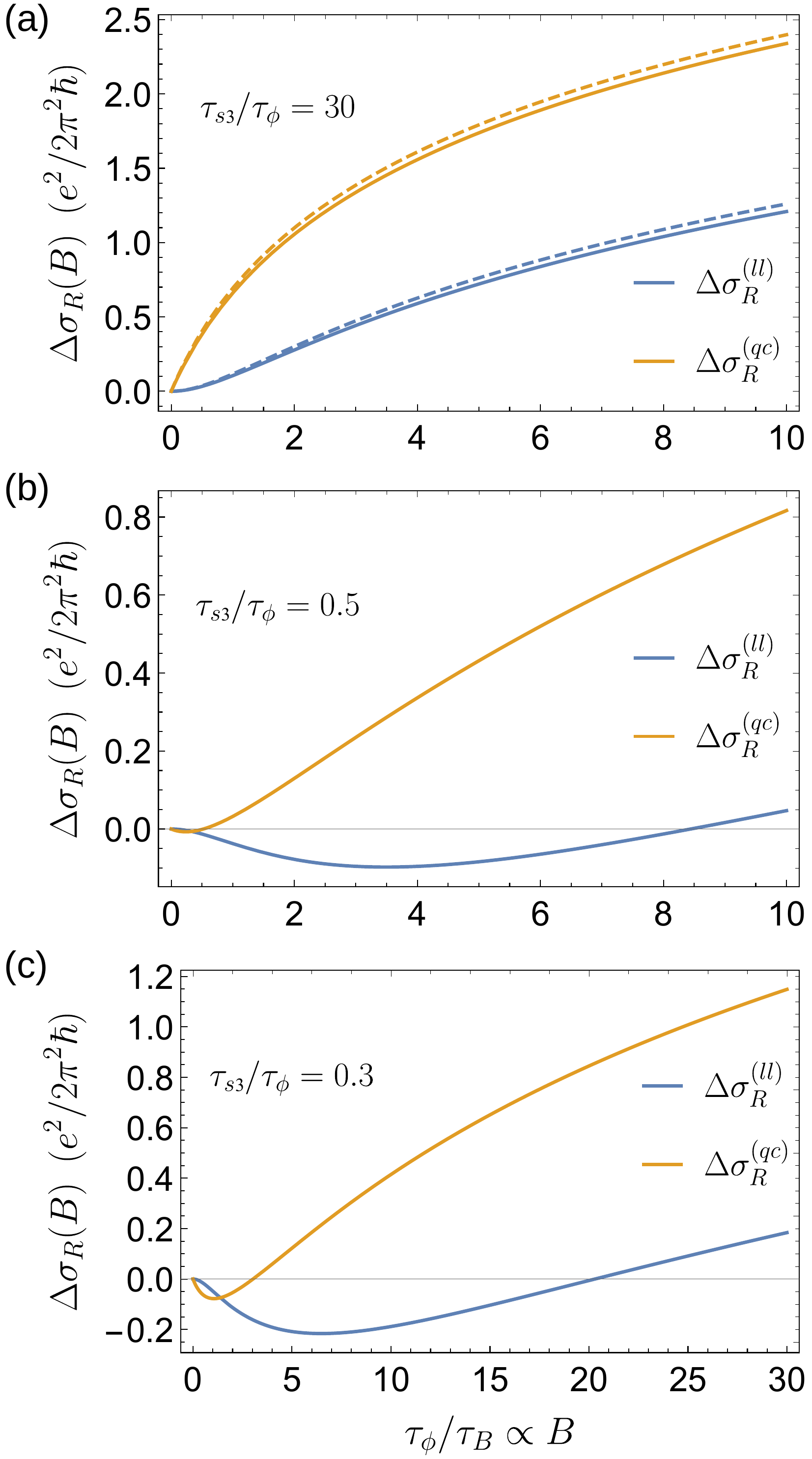}
\caption{Relative magneto-conductivity correction in Landau level (blue) and quasi-classical picture (yellow) obtained by Eqs. (\ref{eq:WAL_qc}) and (\ref{eq:WAL_ll}) in dependence of the ratio of dephasing and magnetic dephasing times $\tau_\phi/\tau_B$ (proportional to the magnetic-field strength $B$) with pure $k$-cubic Dresselhaus SO coupling. 
The ratio of elastic and inelastic scattering times is chosen as $\tau/\tau_\phi=10^{-3}$.
Panels (1)-(3) use different ratios of spin lifetime and inelastic scattering time $\tau_{s3}/\tau_\phi$ as shown in the inset.
  The dashed lines in (a) correspond to the case without SO coupling.}
\label{fig:sigma_R}
\end{figure}
\begin{figure}[t]
\includegraphics[scale=.5]{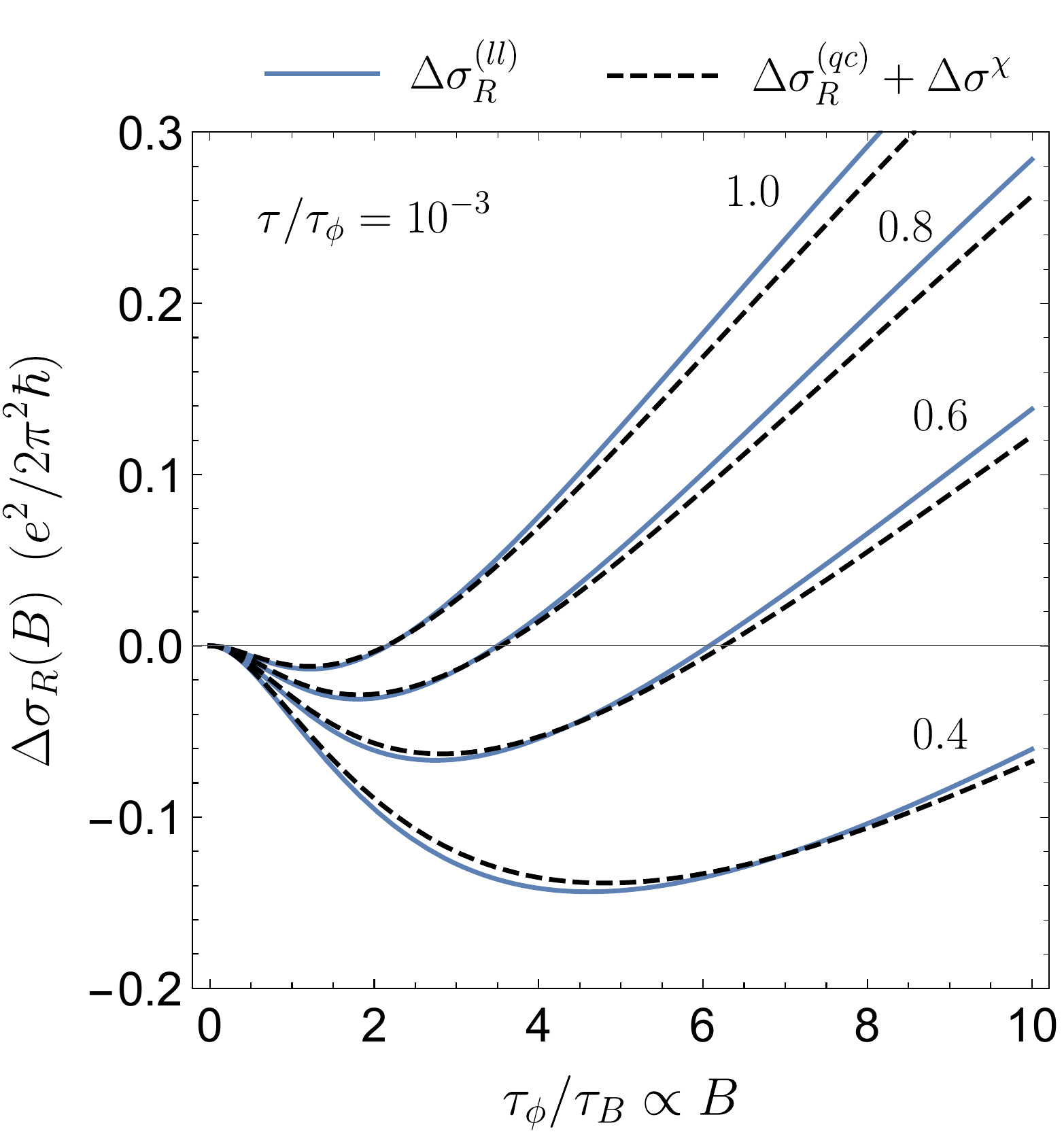}
\caption{Relative magneto-conductivity correction in the Landau-level (blue) resulting from Eq.~(\ref{eq:WAL_ll}) and quasi-classical picture taking into account the first two corrections of the Euler-MacLaurin formula (black dashed) in the diffusive limit, where $\tau\ll\tau_{\phi,B,s3}$, using Eqs.~(\ref{eq:WAL_qc}) and substituting $\nu\rightarrow \nu+f$ with $f$ obtained from Eq.~(\ref{eq:f_approx}).
We select ratio of elastic and inelastic scattering times $\tau/\tau_\phi=10^{-3}$ and vary the ratio of spin lifetime and inelastic scattering time $\tau_{s3}/\tau_\phi$ (the values are displayed next to the corresponding curves).}
\label{fig:sigma_R_SOC_corrections}
\end{figure}

The impact of $k$-cubic Dresselhaus SO coupling on the magneto-conductivity correction was first studied by Altshuler \textit{et al.}\cite{Altshuler1981b} in view of bulk semiconductors but the transfer to 2DEGs is straightforward.\cite{Iordanskii1994,Knap1996}
Although a mere $k$-cubic spin-splitting is not very common in 2DEGs, this is the only case that enables the derivation of an \textit{exact} closed-form expression for the magneto-conductivity correction displaying weak-antilocalization features. 
Since the $k$-cubic Dresselhaus SO coupling yields only $q$-independent terms in the inverse  Cooperon (\ref{eq:Cooperon001}), the Landau levels are decoupled and the calculation follows analogously to the spin-degenerate case of the previous section, apart from an the inclusion of finite spin-relaxation rates in each triplet channel.
Consequently, this example is ideally suited for comparison between Landau-level and quasi-classical approaches in the \textit{weak-antilocalization} regime.

Diagonalizing the operator (\ref{eq:Cooperon001}) for ${\alpha=\beta^{(1)}=0}$, we find the triplet eigenvalues
\begin{align}
E^T_l/\hbar=D_e q^2+\Delta_l,\ \quad l\in\{0,\pm 1\},
\end{align}
where the gaps $\Delta_{\pm 1}=1/\tau_{s3}$ and $\Delta_{0}=2/\tau_{s3}$, with
\begin{align}
1/\tau_{s3}={}&2\tau(k_F\beta^{(3)}/\hbar)^2,
\label{eq:SR_cubicD}
\end{align}
correspond to spin-relaxation rates of homogeneous spin textures with in-plane and out-of-plane spin polarizations, respectively.
The considerations of the previous section remain valid if we replace $1/\tau_\phi\rightarrow1/\tau_\phi+\Delta_l$ with $l\in\{0,\pm 1\}$ in the three triplet channels.
Since the sum over spin indices is now performed explicitly, each singlet and triplet contribution to the conductivity gains an additional prefactor $\pm 1/2$ with a positive sign for each triplet and a negative sign for the singlet channel [cf. Eq.~(\ref{eq:WAL2})].

In analogy with Eqs.~(\ref{eq:WALnoSOC_qc}) and (\ref{eq:WALnoSOC_ll}) the magneto-conductivity corrections in   quasi-classical and Landau-level picture read
\begin{align}
\Delta\sigma^{(qc)}=&{} -\frac{e^2}{4\pi^2\hbar}\left[
\nu\Big(2\tau_{s3}^{-1}\Big)+2\nu\Big(\tau_{s3}^{-1}\Big)-\nu(0)\right]
\label{eq:WAL_qc}
\end{align} 
and 
\begin{align}
\Delta\sigma^{(ll)}=&{} -\frac{e^2}{4\pi^2\hbar}\left[
\xi\Big(2\tau_{s3}^{-1}\Big)+2\xi\Big(\tau_{s3}^{-1}\Big)-\xi(0)\right],
\label{eq:WAL_ll}
\end{align} 
respectively. 
Including the first two Euler-MacLaurin corrections amounts to replacing $\nu\rightarrow\nu+f$ in Eq.~(\ref{eq:WAL_qc}). 
The relative magneto-conductivity obtained by formula (\ref{eq:WAL_ll}) in the diffusive limit is equivalent to expression (39) in Ref.~\onlinecite{Knap1996}.

In Fig.~\ref{fig:sigma_R}, we plot the relative magneto-conductivity correction in Landau-level (blue) and quasi-classical picture (yellow), using the Eqs.~(\ref{eq:WAL_qc}) and (\ref{eq:WAL_ll}), in dependence of the ratio of inelastic scattering and magnetic dephasing times $\tau_\phi/\tau_B$, which is  proportional to the magnetic-field strength $B$, for pure $k$-cubic Dresselhaus SO coupling. 
Panels (a)-(c) use different ratios of $\tau_{s3}/\tau_\phi$ as shown in the insets.
The dashed lines in (a) correspond to the case without SO coupling.
As expected from the observations in the previous section, we see that there is a large discrepancy in the shape of the curves and, in particular, the location of the  minimum, which is essential in experimental data fitting for  an accurate extraction of SO  parameters and spin lifetimes.

Fig.~\ref{fig:sigma_R_SOC_corrections} displays the relative magneto-conductivity correction in the Landau-level approach (blue) together with the quasi-classical approximation but taking into account the first two Euler-MacLaurin corrections (black-dashed) in the diffusive limit, where $\tau\ll\tau_{\phi,B,s3}$.
We select  $\tau/\tau_\phi=10^{-3}$ and vary the ratio $\tau_{s3}/\tau_\phi$, where the values are displayed next to the corresponding curves.
Including the first two Euler-MacLaurin corrections, allows to obtain the magneto-conductivity minimum with high accuracy, contrary to the simplified quasi-classical approach.

\subsection{Rashba or Dresselhaus SO coupling}\label{sec:Iordanski}

The case of pure Dresselhaus SO coupling, explored by Iordanskii \textit{et al.},\cite{Iordanskii1994} is a special one as the rotational symmetry of the spin splitting permits reorganizing the Landau-spin-1 basis $\ket{n}\ket{j,m_j}$  that it block-diagonalizes the Cooperon.
In particular, if we disregard the generally decoupled singlet channel, the basis subsets $\{\ket{n-1}\ket{1,1},\ket{n}\ket{1,0},\ket{n+1}\ket{1,-1}\}$, where $n\geq 1$, split the Cooperon into non-interacting $3\times 3$  blocks that can be diagonalized analytically. 
The remaining states at the bottom of the spectrum $\{\ket{1}\ket{1,-1},\ket{0}\ket{1,0},\ket{0}\ket{1,-1}\}$ have to be included separately as the corresponding basis subsets would contain unphysical states with negative Landau-level quantum number.
Similarly, since the sum runs over Landau levels $n\in[1,N_\text{max}]$, the state $\ket{N_\text{max}+1}\ket{1,-1}$ is taken into account while $\ket{N_\text{max}}\ket{1,1}$ is excluded. 
However, the arising error can be ignored as the contribution of the states $\ket{N_\text{max}}$ to the conductivity is of the order $N_\text{max}^{-1}$ smaller than those at the bottom of the spectrum and thus generally negligible in the diffusive regime, where $N_\text{max}\gg 1$.

Following this recipe, we obtain for the magneto-conductivity correction 
\begin{widetext}
\begin{align}
\Delta\sigma^{(ll)}={}&\frac{e^2}{4\pi^2\hbar}
\bigg\{\xi(0)
-\frac{1}{a_0}
-\frac{\tau_B/2\tau_s+2a_0+1}{a_1(a_0+\tau_B/2\tau_s)-\tau_B/\tau_{s1}}
-\sum_{n=1}^{N_\text{max}}
\frac{3a_n^2+a_n\tau_B/\tau_s-1-(2n+1)\tau_B/\tau_{s1}}{a_{n+1}a_{n-1}(\tau_B/2\tau_s+a_n)-\tau_B/\tau_{s1}[a_n(2n+1)-1]}
\bigg\},
\label{eq:Iordanski1}
\end{align}
\end{widetext}
where $a_n=n+1/2+\tau_B/2\tau_s+\tau_B/2\tau_\phi$.
The spin-relaxation rate $1/\tau_s=1/\tau_{s1}+1/\tau_{s3}$, where 
\begin{align}
1/\tau_{s1,s3}={}&2\tau(k_F\beta^{(1,3)}/\hbar)^2,
\label{eq:SR_linearD}
\end{align}
describes the decay of a spin texture that is homogeneously polarized in the quantum-well plane.
The first term in Eq.~(\ref{eq:Iordanski1}) constitutes the singlet contribution.
The second and third term are due to the separately considered triplet states at the bottom of the spectrum while the last term sums over the remaining reciprocal eigenvalues arising from the decoupled $3\times 3$ triplet blocks.
Without $k$-linear Dresselhaus SO coupling, i.e., for $1/\tau_{s1}=0$, this expression agrees with Eq.~(\ref{eq:WAL_ll}) in the diffusive limit.
Since the Rashba and $k$-linear Dresselhaus Hamiltonians are related by a unitary transformation, the formula~(\ref{eq:Iordanski1}) is also valid for pure Rashba SO coupling (although a different basis subset for block-diagonalization must be used).\cite{Pikus1995}
To obtain the corresponding expression for pure Rashba SO coupling, we must set $\beta^{(3)}=0$, i.e., $\tau_{s}\rightarrow\tau_{s1}$, and replace $\beta^{(1)}$ by the Rashba coefficient $\alpha$.

Owing to Landau quantization, the magneto-conductivity correction (\ref{eq:Iordanski1}) is not a smooth function of $B$.
To remove fluctuations related to the stepwise variation of $N_\text{max}$ with $B$, we can add and subtract in the sum of Eq.~(\ref{eq:Iordanski1}) the term $3/n$, which corresponds to the asymptotic contribution for $n\rightarrow \infty$ of all three triplet channels.
After evaluation of the sum, the added term yields the conductivity shift $-3e^2[\Psi(N_\text{max}+1)+\gamma]/4\pi^2\hbar$, where $\gamma$ denotes the Euler–Mascheroni constant and $N_\text{max}$ can now be formally treated as a continuous quantity.
The subtracted term, on the other hand, allows extending the sum to infinity since both summands approximately cancel each other for $n>N_\text{max}\gg 1$.
If we further employ the asymptotic expansion of the digamma functions, we recover Iordanskii \textit{et al.}'s original expression for the magneto-conductivity correction, Eq.~(13) in Ref.~\onlinecite{Iordanskii1994}.

Even though in the special scenario of either pure Dresselhaus or pure Rashba SO coupling an analytic expression for the magneto-conductivity correction can be obtained, the result still requires a tedious summation over all - in Iordanskii \textit{et al.}'s original result even \textit{infinite} - Landau levels.
Hence, the technical applicability of the formula for experimental data fitting is limited.

\subsection{Cooperon in case of PSH symmetry}\label{sec:PSHsymm}

The particular structure of the Cooperon spectrum in case of PSH symmetry was first pointed out by Pikus \textit{et al.}\cite{Pikus1995}
Setting $\Gamma_3=0$ and $\Gamma_1= 1$ in the operator (\ref{eq:Cooperon001}), the inverse Cooperon can be written as
\begin{align}
\mc{C}^{-1}/D_e\hbar=&{}\;q_x^2+(q_y+Q_0S_x)^2,
\label{eq:Cooperon001_PSH}
\end{align}
and similarly for $\Gamma_1=-1$ if the $x$ and $y$ components of $\mb{q}$ and $\mb{S}$ are interchanged.
In the eigenbasis $\ket{1,m_j'}$ of $S_x$ the three triplet states decouple yielding the eigenvalues
\begin{align}
\mc{C}^{-1}\ket{1,m_j'}/D_e\hbar=&{}\;\left[q_x^2+(q_y+m_j'Q_0)^2\right]\ket{1,m_j'}.
\label{eq:Cooperon001_PSH_eval}
\end{align}
The physical origin for this decoupling is the well-defined spin quantization axis due to the SO-generated effective magnetic field that is collinear in $k$-space on the Fermi circle.
If $\mb{q}$ varies continuously, the spectrum consists of three parabolas where one is centered at $\mb{q}=\mb{0}$ while the other two are displaced by {$\mb{q}=\pm Q_0\mb{\hat{y}}$}.
The minima correspond to the vanishing spin-relaxation rates of the persistent spin textures (\ref{eq:homo}) and (\ref{eq:psh}) for $\eta=0$.

In the computation of the magneto-conductivity, the wave-vector shift in Eq.~(\ref{eq:Cooperon001_PSH_eval}) can be neglected since the commutation relations $[q_x,q_y]$ do not change when $q_y$ is displaced by $m_j'Q_0$ and the modifications in the upper cut-off are irrelevant in the diffusive regime, where $Q_0\ll 1/\sqrt{D_e\tau}$.
The latter represents the usual precondition for the D'yakonov-Perel' mechanism, which requires that the spin-precession length should be much longer than the mean free path.
As result of the absence of $q$-linear terms, the Landau levels are decoupled and magneto-conductivity correction can be analogously calculated as in the case of pure $k$-cubic Dresselhaus SO coupling (cf. Sec.~\ref{sec:pureCubicDress}).
Since the Cooperon spectrum is gap-less, the inferred expression for the magneto-conductivity correction resembles that of a system without SO coupling and exhibts weak-localization characteristics.

Initially, such situation appears to be not very useful as for experimentally extracting SO-related parameters the weak-\textit{anti}localization features with the typical local extrema in the negative magneto-conductivity are essential.
However, we can use the exact solvability at the PSH-symmetry as a basis for a perturbative expansion of the Cooperon spectrum, providing analogous structure but including small spin-relaxation gaps due to the finite lifetime of the long-lived spin textures.
The resulting magneto-conductivity expression turns out to be a good approximation even in a parameter regime where a transition to \textit{weak-antilocalization} has already taken place and is therefore very helpful for experimental data fitting.

\section{General 2DEG near PSH symmetry}\label{sec:SOCaab}

It was demonstrated in Ref.~\onlinecite{Kammermeier2016PRL} that a PSH symmetry generally exists in 2DEGs if the quantum-well growth direction has at least two Miller indices equal in modulus, the Rashba and $k$-linear Dresselhaus coefficients $\alpha$ and $\beta^{(1)}$ are suitably matched, and the $k$-cubic Dresselhaus coefficient $\beta^{(3)}$ vanishes.
Small deviations from that symmetry either due to a misalignment of Rashba and $k$-linear Dresselhaus coefficients or due to the presence of finite $k$-cubic Dresselhaus SO coupling induce finite spin-relaxation rates  $1/\tau_{\rm homo}$ and $1/\tau_{\rm PSH}$  [Eqs.~(\ref{eq:HOMOrate}) and (\ref{eq:PSHrate})] of the otherwise persistent homogeneous and helical spin textures (cf. Sec.~\ref{sec:SDE}).

As shown in Eq.~(\ref{eq:SDspectrumPSH}), near the PSH-symmetry point the spectrum of the spin-diffusion operator (and therewith the spectrum of the inverse Cooperon) can be approximated by three parabolas analogously to expression~(\ref{eq:Cooperon001_PSH_eval}) but  with gapped minima determined by the finite spin-relaxation rates of the long-lived spin textures.
Following the reasoning of the previous section, we can neglect the wave-vector shift $\mb{Q}$ and write the eigenvalues of the triplet sector of the inverse Cooperon for a 2DEG grown along unit vector $\mb{\hat{n}}$ as
\begin{align}
E^T_l/\hbar\approx D_e q^2+\tilde{\Delta}_l,\ \quad l\in\{0,\pm 1\},
\end{align}
with the gaps $\tilde{\Delta}_{\pm 1}=1/\tau_{\rm PSH}$ and $\tilde{\Delta}_{0}=1/\tau_{\rm homo}$.
The simplified Cooperon spectrum now allows to compute a closed-form expression for the magneto-conductivity correction that reads in the Landau-level picture
\begin{align}
\Delta\sigma^{(ll)}&={}\frac{e^2}{4\pi^2\hbar}\bigg[\xi(0)-\xi\Big(\tau_\text{homo}^{-1}\Big)-2\xi\Big(\tau_\text{PSH}^{-1}\Big)
\bigg].
\label{eq:WAL_ll_general}
\end{align} 
Using the approximation (\ref{eq:xi_approx}) for $\xi$  in the diffusive limit ($\tau\ll\tau_{\phi,B,\text{homo,PSH}}$) and considering a [001] quantum-well growth direction ($\eta=0$), Eq.~(\ref{eq:WAL_ll_general}) is equivalent to the one recently obtained by Weigele \textit{et al.} in Ref.~\onlinecite{Weigele2020}.

As analogously shown in Sec.~\ref{sec:pureCubicDress}, we may alternatively use the quasi-classical approximation and include the first two Euler-MacLaurin corrections, which amounts to substituting $\xi\rightarrow\nu+f$ in Eq.~(\ref{eq:WAL_ll_general}).
Even though the use of the improved quasi-classical expression seems not to offer a practical advantage for fitting, the function $\nu+f$ unlike $\xi$ does not contain any special functions,  which can be useful for analytic purposes. 
(The function $\xi$ depends on the Digamma function, which - being the logarithmic derivative of the Gamma function - is strictly speaking not a closed-form expression.)

In the following, we give examples to validate the applicability of Eq.~(\ref{eq:WAL_ll_general}) and compare with previous models.

\begin{figure*}[t]
\includegraphics[scale=.52]{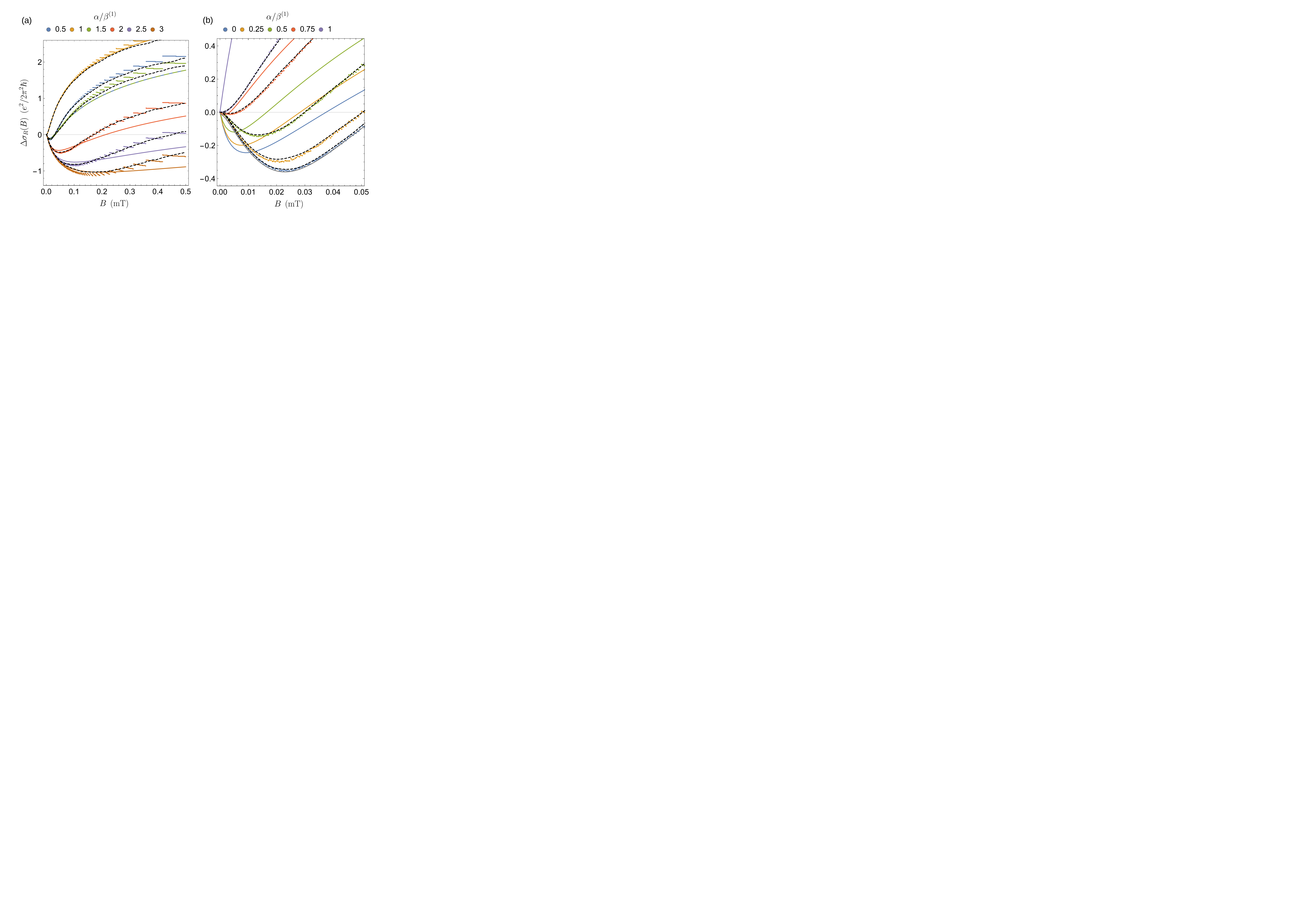}
\caption{Relative magneto-conductivity correction for a GaAs (001)-2DEG corresponding to the experimental setup in Ref.~\onlinecite{Saito2021} with elastic (inelastic) scattering time $\tau=\SI{1.53}{ps}$ ($\tau_\phi=\SI{2.33}{ns}$), effective mass $m=0.067 m_e$ with electron mass $m_e$, electron density ${n_e =\SI{3.0e15}{m^{-2}}}$, and $k$-linear and $k$-cubic Dresselhaus SO  coefficients $ {\beta^{(1)} =\SI{1.9}{meV \AA}}$ and $ {\beta^{(3)} =\SI{0.4}{meV \AA}}$, respectively.
The ratio of Rashba and $k$-linear Dresselhaus coefficients $\alpha$ and $\beta^{(1)}$ is varied in the range ${[0.5, 3]}$ in (a) and ${[0, 1]}$ in (b) as indicated by the distinct colors in the plot legends.
The colored discontinuous lines in (a) and dots in (b) correspond to full numerical calculation of $\Delta\sigma_R$ by diagonalization of the operator (\ref{eq:Cooperon001}) in the Landau-spin-1 basis.
The black dashed lines display the result obtained by a Monte-Carlo-based real-space simulation as developed by Sawada \textit{et al.} in Ref.~\onlinecite{Sawada2017}, which also applies beyond the diffusive regime. The numerical data is compared with the closed-form expression (\ref{eq:WAL_ll_general}) in (a) and with Eq.~(19) of Ref.~\onlinecite{Marinescu2019} derived by Marinescu \textit{et al.} in (b) (colored solid lines).
The additional gray solid line in (b) is computed by Iordanski \textit{et al.}'s analytic formula derived in Ref.~\onlinecite{Iordanskii1994} and reviewed in Sec.~\ref{sec:Iordanski}.}
\label{fig:sigma_R_001_Saito}
\end{figure*}

\begin{figure*}[t]
\includegraphics[scale=.52]{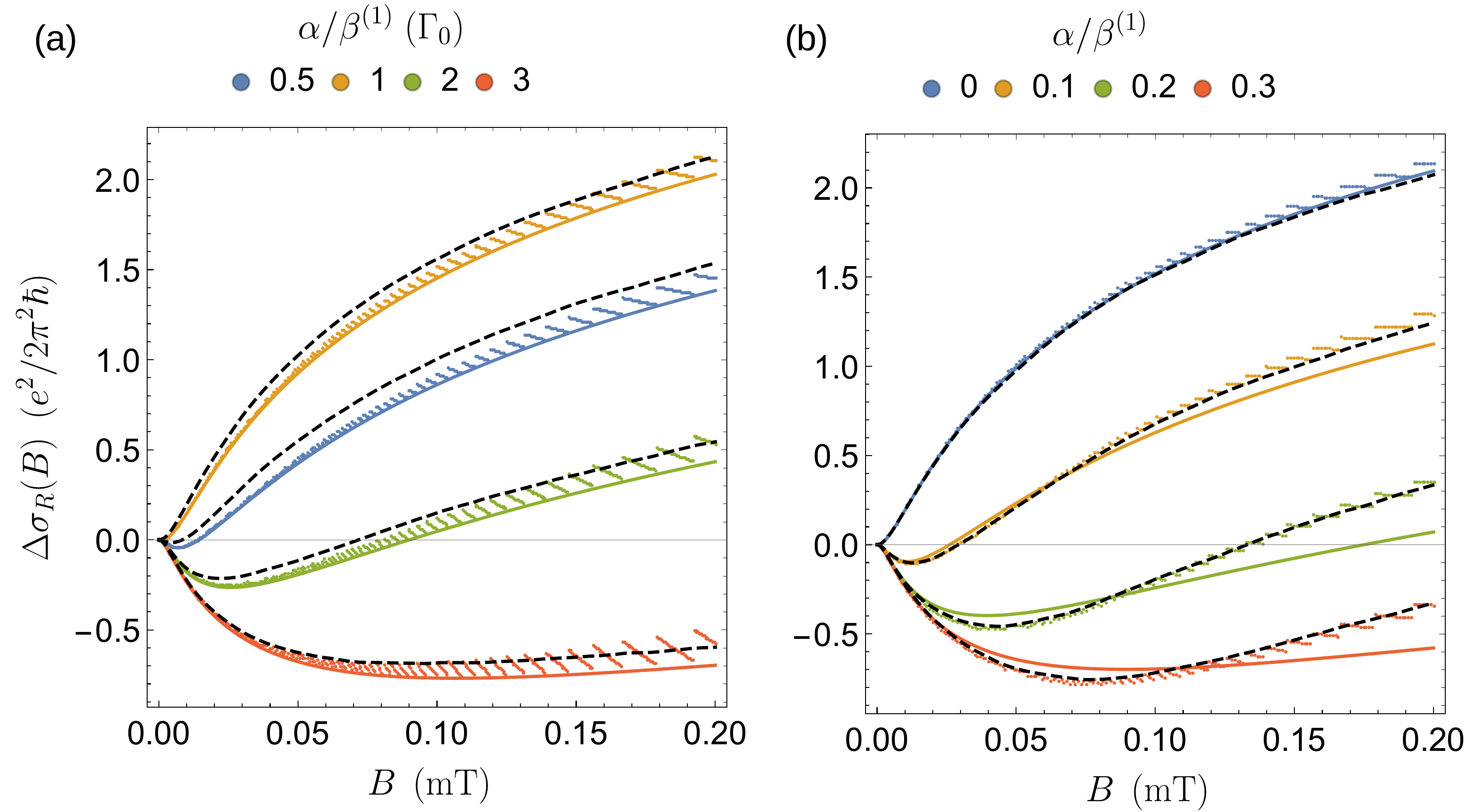}
\caption{Relative magneto-conductivity correction for a hypothetical GaAs 2DEG with a quantum-well growth direction along [113] in (a) and [110]  in (b). 
As system parameters, we selected  the elastic (inelastic) scattering time $\tau=\SI{1.53}{ps}$ ($\tau_\phi=\SI{2.33}{ns}$), effective mass $m=0.067 m_e$ with electron mass $m_e$, electron density ${n_e =\SI{3.0e15}{m^{-2}}}$, and $k$-linear and $k$-cubic Dresselhaus SO coefficients $ {\beta^{(1)} =\SI{8.8}{meV \AA}}$ and $ {\beta^{(3)} =\SI{0.4}{meV \AA}}$, respectively.
The ratio of Rashba and $k$-linear Dresselhaus coefficients $\alpha$ and $\beta^{(1)}$ is varied in the range ${[0.5, 3]}$ in (a) and ${[0, 0.3]}$ in (b) as indicated by the distinct colors in the plot legends.
The colored dots correspond to full numerical calculation of $\Delta\sigma_R$ using the operator (\ref{eq:sdo}) for (a) $\eta=1/\sqrt{11}$ and (b) $\eta=1/\sqrt{2}$  and taking into account Landau quantization.
The black dashed lines display the result obtained by a Monte-Carlo-based real-space simulation as developed by Sawada \textit{et al.} in Ref.~\onlinecite{Sawada2017}. 
The colored solid lines represent the magneto-conductivity correction described by the closed-form expression (\ref{eq:WAL_ll_general}).}
\label{fig:sigma_R_113_110}
\end{figure*}

\subsection{Model comparison for a realistic (001)-2DEG}
\subsubsection{Weakly broken PSH symmetry}

In Ref.~\onlinecite{Saito2021}, the experimentally extracted magnetoconductance profile of a (001)-2DEG in a GaAs/AlGaAs heterostructure was fitted by the approximate magneto-conductivity expression (\ref{eq:WAL_ll_general}) for parameters in the vicinity of the PSH symmetry.
Thereby, the authors achieved an electrical evaluation of the spin-relaxation rates of both homogeneous and helical long-lived spin textures simultaneously.
Note that this constitutes an advantage over, e.g., optical spin-orientation measurements, where only the helical spin-lifetime is accessible.\cite{Kohda2017} 
The accuracy of the obtained rates has been independently verified  by inspection of the weak-localization anisotropy in wire geometries under application of an in-plane magnetic field and extracting SO coefficients following the method of Refs.~\onlinecite{Sasaki2014,Nishimura2021}.

To prove the accuracy and usefulness of Eq.~(\ref{eq:WAL_ll_general}), we show that for the device parameters in line with the investigations in Ref.~\onlinecite{Saito2021}, the formula (\ref{eq:WAL_ll_general}) agrees well with the full numerical calculation by diagonalizing the operator (\ref{eq:Cooperon001}) in the Landau-spin-1 basis.\footnote{To confirm the correctness of our numerical calculations, we reproduced Figs.~1 and 2 of Pikus \textit{et al.} in Ref.~\onlinecite{Pikus1995}, who performed analogous numerical calculations but used the extension of the summation $N_{\rm max }\rightarrow \infty$ equivalently to Iordanski \textit{et al.}'s approach  to smoothen their curves (cf. Sec.~\ref{sec:Iordanski}).}
For additional validation, we compare also with the results obtained by a Monte-Carlo-based real-space simulation as introduced by Sawada~\textit{et al.} in Ref.~\onlinecite{Sawada2017}.
Here, the conductivity correction is determined by considering the phase-breaking effect of magnetic field and SO coupling on the return probability of electrons propagating along predetermined closed-loop trajectories.
As this method does not set any precondition on the relation of scattering and phase-breaking time scales, it is also valid beyond diffusion approximation.

In the following, we assume an elastic scattering time $\tau=\SI{1.53}{ps}$, inelastic scattering time $\tau_\phi=\SI{2.33}{ns}$, effective mass $m=0.067 m_e$ of GaAs with bare electron mass $m_e$, electron sheet density ${n_s =\SI{3.0e15}{m^{-2}}}$, $k$-linear and $k$-cubic Dresselhaus SO coefficients $ {\beta^{(1)} =\SI{1.9}{meV \AA}}$ and $ {\beta^{(3)} =\SI{0.4}{meV \AA}}$, respectively.\footnote{Note that Ref.~\onlinecite{Saito2021} defines the $k$-linear and $k$-cubic Dresselhaus coefficient as $\beta_1$ and $\beta_3$, respectively, which are related by $\beta^{(1)}=\beta_1-\beta_3$ and $\beta^{(3)}=\beta_3$  to the definitions in the present paper.}
In the experiment, the $k$-cubic Dresselhaus SO coupling breaks the PSH symmetry only weakly such that the weak-localization features are still observable for the PSH condition ${\Gamma_1=1}$ for Rashba and $k$-linear Dresselhaus coefficients.
A crossover from weak localization to weak antilocalization is however achieved by an additional variation of the ratio $\Gamma_1$ in the range $[0.5,3]$.

In Fig.~\ref{fig:sigma_R_001_Saito}(a), we display the resulting relative magneto-conductivity correction obtained by the approximate
closed-form expression (\ref{eq:WAL_ll_general}) near the PSH regime (smooth colored solid lines), the exact numerical calculation in the Landau-spin-1 basis (discontinuous colored solid lines), and the Monte-Carlo-based real-space simulation\cite{Sawada2017} (black dashed lines).
The different colors correspond to distinct ratios $\Gamma_1$ as indicated by the plot legend.
The numerical data from diagonalizing the multiband Cooperon exhibits jumps due to the step-wise change of the maximum Landau-level $N_\text{max}$ with $B$, where the width of steps becomes wider with increasing magnetic field. 
We see that all three approaches show very good agreement. 
In particular, the characteristic magneto-conductivity minima that are essential for a reliable spin-lifetime extraction are well reproduced.
Hence, the utilization of expression (\ref{eq:WAL_ll_general}) for fitting experimental data  as done in Ref.~\onlinecite{Saito2021} is indeed suitable.

\subsubsection{General spin-orbit parameter configurations}

For SO parameter configurations far from PSH symmetry, the expression (\ref{eq:WAL_ll_general}) is no longer valid.
In the recent paper of Marinescu \textit{et al.}\cite{Marinescu2019}, a  magneto-conductivity  model for \textit{arbitrary} ratios of Rashba and Dresselhaus SO coupling was presented. 
The derivation involves, on the one hand, approximations of the Cooperon spectrum wrt to the SO coupling, on the other hand, a quasi-classical treatment of the magnetic field with the assumption of  quasi-continuous Cooperon bands.  
While these simplifications enabled an analytical integration over wave-vector magnitude $q$ [cf. Eq.~(\ref{eq:WAL4})], the final result further necessitates the evaluation of an angular integral and is thus not in closed form.
The authors used a successful comparison with experiment for model validation.

Here, we test the accuracy of Marinescu \textit{et al.}'s magneto-conductivity model, Eq.~(19) in Ref.~\onlinecite{Marinescu2019}, in a GaAs (001)-2DEG with the same parameter configurations as above but with the ratio $\Gamma_1$ in the range $[0,1]$.
This regime includes the special case of pure Dresselhaus SO coupling ($\Gamma_1=0$),  which is far from PSH symmetry and allows for the particular analytic expression  (\ref{eq:Iordanski1}), yielding the well-known formula of Iordanskii~\textit{et al}.\cite{Iordanskii1994}
In Fig.~\ref{fig:sigma_R_001_Saito}(b) we compare the relative magneto-conductivity correction computed by Marinescu \textit{et al.}'s formula (colored solid lines),\footnote{To ensure the correct implementation of Marinescu \textit{et al.}'s formula, we numerically reproduced Fig.~2 in Ref.~\onlinecite{Marinescu2019}.} the full numerical calculation by diagonalization of operator (\ref{eq:Cooperon001}) in the Landau-spin-1 basis (colored dotted lines), the Monte-Carlo-based real-space simulation\cite{Sawada2017} (black dashed lines), and Iordankii~\textit{et al.}'s analytic result, Eq.~(13) in Ref.~\onlinecite{Iordanskii1994}, (gray solid line).
Both numerical approaches and the result by Iordanskii \textit{et al.}, in the case of pure Dresselhaus SO coupling, show perfect agreements.
On the contrary, we note a significant discrepancy with Marinescu \textit{et al.}'s formula for SO parameters far as well as close to PSH symmetry.
Similar disagreement can be found when using the magneto-conductivity expression of Kammermeier~\textit{et al.} in Ref.~\onlinecite{Kammermeier2016PRL}, limited to $k$-linear SO terms and the vicinity of the PSH regime, because it also employs the quasi-classical approximation (here not shown). 

In summary, among the available analytic expressions for the magneto-conductivity correction in (001)-2DEGs there are only two models that enable accurate data extraction from experiment.
Formula (\ref{eq:WAL_ll_general}) can be used close to the PSH symmetry, where Rashba and effective $k$-linear Dresselhaus coefficients are of similar order, and formula (\ref{eq:Iordanski1}), or the ensuing expression of Iordanskii \textit{et al}.,\cite{Iordanskii1994} when either Rashba or Dresselhaus SO coupling is negligible.

\subsection{Exemplary 2DEGs grown along [113] and [110]}

To underline the applicability to other growth directions and encourage further experimental studies, we demonstrate now the results for two exemplary (113)- and (110)-2DEGs, which are particularly interesting for the following reasons.

In Ref.~\onlinecite{Iizasa2020}, it was theoretically predicted that the PSH-lifetime limitation due to the $k$-cubic Dresselhaus coefficient is minimized in quantum wells with a [225] low-symmetry growth direction. 
Although such quantum wells are rather unusual, a [225] growth direction can be approximated by the [113] growth direction, which has been realized in experiment and according to the results in Ref.~\onlinecite{Iizasa2020} should yield a similarly strong PSH-lifetime enhancement.
For instance, quantum wells grown along [113] have been reported to exhibit long hole-spin-relaxation times\cite{Ganichev2002,Schneider2004} and recently gained notable attention due to large hole g-factor anisotropy.\cite{Gradl2014,Gradl2018}
The PSH-symmetry point for a (113)-2DEG corresponds to a small ratio of Rashba and $k$-linear Dresselhaus coefficients $\Gamma_0(\eta=1/\sqrt{11})\approx 0.16$.

In a (110)-2DEG, the PSH symmetry is generated in the absence of Rashba SO coupling, that is, ${\Gamma_0(\eta=1/\sqrt{2})= 0}$.
The resulting uniaxial SO field has only components perpendicular to the quantum-well plane.
Spin lifetimes in (110)-2DEGs have been experimentally studied by means of optical spin-orientation measurements.\cite{Ohno1999,Mueller2008,Chen2014a} 
Since optically the spins are typically polarized perpendicular to the quantum-well, only the homogeneous persistent spin textures can be adressed in these systems.
Probing lifetime and dynamics of a PSH would, however, require a spin excitation parallel to the quantum-well plane as the PSH precesses about the axis of the SO field, which is experimentally difficult to achieve.
Chen \textit{et al.} circumvented this issue by manipulating the spin-quantization axis through the application of a tilted magnetic field with in-plane and out-of-plane components and performing space and time-resolved Kerr-rotation measurements.\cite{Chen2014a}
Using the manifestation of the PSH lifetime in the magneto-conductivity correction provides an alternative access that does not necessitate such interventions and motivates an electrical-evaluation approach of the spin relaxation as presented in Ref.~\onlinecite{Saito2021}.

In Fig.~\ref{fig:sigma_R_113_110}, we display the relative magnetoconductivity correction  for a hypothetical GasAs  2DEG with growth directions along [113] in  (a) and [110] in (b).
We employ analogous system parameters as in the previous section but for illustrative purposes slightly elevated $k$-linear Dresselhaus coefficient $\beta^{(1)}=\SI{8.8}{meV\AA}$.
The ratio $\Gamma_1$ is varied around the respective PSH condition, where we selected the range [0.5, 3] in (a) and [0, 0.3] in (b).
Similar to the (001)-2DEG, we find in both cases good agreement between the numerical diagonalization of the Cooperon, related to the  spin-diffusion operator (\ref{eq:sdo}), in the Landau-spin-1 basis (colored dotted lines), the Monte-Carlo-based real-space simulation by Sawada \textit{et al.}\cite{Sawada2017} (black dashed lines), and the closed-form expression (\ref{eq:WAL_ll_general}) shown by the colored solid lines.

\section{Conclusion}\label{sec:concl}

We reviewed the applicability of models for the quantum-interference correction to the magneto-conductivity with both Rashba and Dresselhaus SO coupling.
In particular, we looked at the recent model of Marinescu~\textit{et al.},\cite{Marinescu2019} which aims to solve the long-standing problem of a missing closed-form magneto-conductivity expression for \textit{arbitrary} ratios of Rashba and Dresselhaus SO coefficients.
It is shown that this model is unsuitable for experimental parameter fitting due to a significant lack of accuracy that results from a quasi-classical treatment of the phase-breaking mechanism of the magnetic field.
The discrepancy between the quasi-classical approximation and the  accurate Landau-level approach is related to the replacement of a sum over Landau levels by an integral over wave-vectors.
In some configurations, the error can be reduced by including the first two corrections given by the Euler-MacLaurin formula for integral approximation of a sum.

However, this remedy works only in parameter regimes where Landau levels decouple and an approximate solution - including Landau quantization - is also available.
Such a scenario is given by the PSH regime, where persistent spin textures emerge and simultaneously a transition between weak localization and weak antilocalization occurs.
In this special case, an exact solution for the magneto-conductivity correction including Landau quantization exits.\cite{Pikus1995}
This exact solvability can be used as a basis to derive an approximate closed-form description of the magneto-conductivity correction near the PSH symmetry, as shown recently by Weigele 
\textit{et al.}\cite{Weigele2020} for a (001)-2DEG.
Combing the recent findings,\cite{Kammermeier2016PRL,Iizasa2020} we demonstrate that this method can be generalized to 2DEGs with at least two growth-direction Miller indices equal in modulus.
The resulting expression is a function of spin lifetimes of the long-lived spin textures and thus particularly attractive for exploring the possibility of long-lasting spin coherence.

The validity of the magneto-conductivity formula near the PSH regime is explicitly demonstrated by comparing with the results from distinct numerical calculations in (001)-, (113)-, and (110)-2DEGs.
The simplicity of the derived expressions enable a practical and accurate fitting of experimental data in a wide range of 2DEGs and a broad parameter regime.
For \textit{arbitrary} Rashba and Dresselhaus SO coefficients, however, a numerical approach is inevitable for a reliable description of the magneto-conductivity correction due to weak antilocalization.

\section{Acknowledgment}

M.Ka. and U.Z. acknowledge the support by the Marsden Fund Council from New Zealand government funding (contract no.\ VUW1713), managed by the Royal Society Te Ap\={a}rangi.
M.Ka. was further supported by the Center for Science and Innovation in Spintronics at Tohoku University, Japan.
M.K. is supported via Grant-in-Aid for Scientific Research (Grants No. 21H04647) by the Japan Society for the Promotion of Science. This work was also supported by JST FOREST Program (Grant Number JPMJFR203C). 
We would like to thank Prof.~J. Nitta and P. Wenk for the fruitful discussions.

\appendix

\section{Spin-diffusion operator}\label{app:sdo}

In a zinc-blende 2DEG with Rashba and Dresselhaus SO coupling and general growth direction $\hat{\mb{n}}=(\eta,\eta,\sqrt{1-2\eta^2})$ with $\eta\in[0,1/\sqrt{2}]$, the spin-diffusion operator $\sdo(\mb{q})$ in Fourier space, entering the spin-diffusion equation (\ref{eq:sde}), reads as\cite{Iizasa2020} 
\begin{equation}
\sdo = 
\begin{pmatrix}
\tilde{\Lambda}_{xx}   &  \tilde{\Lambda}_{xy}  & \tilde{\Lambda}_{xz}\\
\tilde{\Lambda}_{xy}^\ast   &  \tilde{\Lambda}_{yy}  & \tilde{\Lambda}_{yz}\\
\tilde{\Lambda}_{xz}^\ast  &  \tilde{\Lambda}_{yz}^\ast  & \tilde{\Lambda}_{zz}\\
\end{pmatrix}
,
\label{eq:sdo}
\end{equation}
with 
\begin{align}
\tilde{\Lambda}_{xx}=&\cfrac{4\pi^2}{\tau_0}\Bigg[\cfrac{1 - 18 \eta^2 + 105 \eta^4 - 144 \eta^6}{4}- \cfrac{ (1-9 \eta^2)n_z}{2} \Gamma_1\nonumber\\
&+\cfrac{\Gamma_1^2}{4} + \cfrac{1 + 10 \eta^2 - 15 \eta^4}{4} \Gamma_3^2 +\cfrac{\mb{q}^2}{Q_0^2}\Bigg],\\
\tilde{\Lambda}_{xy}=&\cfrac{4\pi^2}{\tau_0}\,i \sqrt{2}  (\eta - 3 \eta^3) \cfrac{q_y}{Q_0},\\
\tilde{\Lambda}_{xz}=&\cfrac{4\pi^2}{\tau_0}\Bigg\{\cfrac{\sqrt{2}(\eta-9\eta^5)n_z}{4}+
\left[  \cfrac{\sqrt{2}(\eta-3\eta^3)}{4} -i\cfrac{q_x}{Q_0} \right]\Gamma_1\nonumber\\
&-\cfrac{3(\eta-4\eta^3+3\eta^5)n_z}{2\sqrt{2}}\Gamma_3^2-i(1-9\eta^2)n_z\cfrac{q_x}{Q_0}\Bigg\},\\
\tilde{\Lambda}_{yy}=&\cfrac{4\pi^2}{\tau_0}\,\Bigg[\cfrac{1 + 6 \eta^2 - 15 \eta^4}{4}+\cfrac{(1+3\eta^2)n_z}{2}\Gamma_1+\cfrac{\Gamma_1^2}{4}\nonumber\\
&+\cfrac{1 + 10 \eta^2 - 15 \eta^4}{4}\Gamma_3^2+ \cfrac{\mb{q}^2}{Q_0^2}\Bigg],\\
\tilde{\Lambda}_{yz}=&-\cfrac{4\pi^2}{\tau_0}\,i(\Gamma_1 + n_z + 3 \eta^2 n_z)  \cfrac{q_y}{Q_0},\\
\tilde{\Lambda}_{zz}=&\cfrac{4\pi^2}{\tau_0}\Bigg[\cfrac{1 - 8 \eta^2 + 57 \eta^4 - 90 \eta^6}{2}+6 \eta^2 n_z\Gamma_1 +\cfrac{\Gamma_1^2}{2}\nonumber\\
&+\cfrac{1 - 8 \eta^2 + 21 \eta^4 - 18 \eta^6}{2}\Gamma_3^2+ \cfrac{\mb{q}^2}{Q_0^2}\Bigg],
\end{align}
where we used the basis vectors $\hat{\mb{x}}=(n_z,n_z,-2\eta)/\sqrt{2}$, $\hat{\mb{y}}=(-1,1,0)/\sqrt{2}$, and $\hat{\mb{z}}=\hat{\mb{n}}$.
The parameter $1/\tau_0=D_eQ_0^2/4\pi^2$ corresponds to the spin-precession rate of the persistent spin helix (\ref{eq:psh}) along the $\hat{\mb{y}}$-axis in a (001)-2DEG with spin-helix  wave vector $Q_0=4m\beta^{(1)}/\hbar^2$ and 2D diffusion constant $D_e$.
The underlying Hamiltonian is given in Eq.~(\ref{eq:Hamiltonian}).
If anisotropic scattering is prevalent, we need to distinguish elastic scattering times $\tau_{1,3}$ related to first and third angular harmonics in the SO terms, which amounts to replacing $\tau\rightarrow \tau_1$ and $\beta^{(3)}\rightarrow \beta^{(3)}\sqrt{\tau_3/\tau_1}$.

\section{Relation between spin-diffusion operator and Cooperon} \label{app:SDO_Cooperon} 

In Ref.~\onlinecite{Wenk2010} it was shown that there exists a unitary transformation $U$ that connects the triplet sector of the inverse Cooperon $\mc{C}^{-1}$ with the spin-diffusion operator $\sdo$.
The transformation is given by $
\mc{C}^{-1}= \hbar\,U \sdo U^\dag$, where
\begin{align}
U&={}\begin{pmatrix}
-1&i&0\\
0&0&\sqrt{2}\\
1&i&0
\end{pmatrix}/\sqrt{2}
\end{align}
links the components of the spin-density $\mathbf{s}=(s_x,s_y,s_z)^\top$ with the triplet basis $\mathbf{\tilde{s}}=(\ket{1,1},\ket{1,0},\ket{1,-1})^\top$   via $\mathbf{\tilde{s}}={}U\,\mathbf{s}$.

\section{The Euler-MacLaurin formula}\label{app:sum_integral}

The differences between sum and integral can be computed by the Euler-Maclaurin formula. 
It states that for  integers $N,\,p\geq 1$ and for any function $g\in C^p[a,b]$ we have
\begin{align}
w\sum_{n=0}^N &g(a+n\, w)-\int_a^b du\;g(u)={}\frac{w}{2}[g(a)+g(b)]\notag\\
&+\sum_{j=2}^p w^j\frac{B_j}{j!}[g^{(j-1)}(u)]_a^b+r_p(a,b,N),
\label{eq:EML_formula}
\end{align}
where $B_j$ are Bernoulli coefficients, $w=(b-a)/N$ the segment width, and $r_p(a,b,N)$ is the remainder
\begin{align}
r_p(a,b,N)={}&-\frac{w^p}{p!}\int_a^bdu\;P_p\left(\frac{a-u}{w}\right)g^{(p)}(u),
\end{align}
where $P_p$ is the $p$-th Bernoulli periodic function.~\cite{EulerMacLaurinFormula}
(Notice that $B_3=B_5=B_7=...=0$.)
The first term on the rhs of Eq.~(\ref{eq:EML_formula}) can be obtained by approximating the integral by a sum using the trapezoidal rule with $N+1$ $w$-equidistant  evaluation points.
The other terms may be viewed as an extension of the trapezoidal rule by including corrections arising from Taylor expanding the integrand at the evaluation points~\cite{DevriesBook1994}.
Notice that the asymptotic expansion does usually not converge and past certain values of $p$ the terms increase rapidly.

In the main text, we will limit ourselves to the inclusion of the first two, still well-behaving, corrections for a unit grid $w=1$ on the rhs of Eq.~(\ref{eq:EML_formula}) and neglect the remainder $r_p$.
As relevant summation elements, we identify 
\begin{align}
g(n)=-\frac{e^2}{2\pi^2\hbar}\left[ n+\frac{\tau_B}{2\tau_\phi'(x)}\right]^{-1},
\label{eq:g}
\end{align}
where the integration and summation limits correspond to $a=0$ and $b=N=N_\text{max}$.
Using Eq.~(\ref{eq:g}), we obtain the leading two Euler-Maclaurin corrections
\begin{align}
\Delta\sigma^{\chi_{1}}={}&\frac{1}{2}[g(0)+g(N_\text{max})]\label{eq:correction1},\\
\Delta\sigma^{\chi_{2}}={}&\frac{1}{12}[g'(N_\text{max})-g'(0)]
\label{eq:correction2},
\end{align}
to be added to the quasi-classical magneto-conductivity expression $\Delta\sigma^{(qc)}$.

\section{Spin-1 basis matrices}\label{app:spin_matrices}

The basis matrices for the triplet sector  of a total {spin-1} system  read as
\begin{align}
S_x&={}
\,\frac{1}{\sqrt{2}}
\begin{pmatrix}
0&1&0\\
1&0&1\\
0&1&0
\end{pmatrix},\quad 
S_y={}
\,\frac{i}{\sqrt{2}}
\begin{pmatrix}
0&-1&0\\
1&0&-1\\
0&1&0
\end{pmatrix},\notag\\ 
 S_z&={}
\,
\begin{pmatrix}
1&0&0\\
0&0&0\\
0&0&-1
\end{pmatrix},
\label{eq:spin_matrices}
\end{align}
in the triplet-basis $\ket{j=1,m_j\in\{0, \pm 1\}}$ with the order $\{\ket{1,1},\ket{1,0},\ket{1,-1}\}$.

\bibliographystyle{apsrev4-1}

\bibliography{MK}

\def\url#1{}
\begin{thebibliography}{68}%
\makeatletter
\providecommand \@ifxundefined [1]{%
 \@ifx{#1\undefined}
}%
\providecommand \@ifnum [1]{%
 \ifnum #1\expandafter \@firstoftwo
 \else \expandafter \@secondoftwo
 \fi
}%
\providecommand \@ifx [1]{%
 \ifx #1\expandafter \@firstoftwo
 \else \expandafter \@secondoftwo
 \fi
}%
\providecommand \natexlab [1]{#1}%
\providecommand \enquote  [1]{``#1''}%
\providecommand \bibnamefont  [1]{#1}%
\providecommand \bibfnamefont [1]{#1}%
\providecommand \citenamefont [1]{#1}%
\providecommand \href@noop [0]{\@secondoftwo}%
\providecommand \href [0]{\begingroup \@sanitize@url \@href}%
\providecommand \@href[1]{\@@startlink{#1}\@@href}%
\providecommand \@@href[1]{\endgroup#1\@@endlink}%
\providecommand \@sanitize@url [0]{\catcode `\\12\catcode `\$12\catcode
  `\&12\catcode `\#12\catcode `\^12\catcode `\_12\catcode `\%12\relax}%
\providecommand \@@startlink[1]{}%
\providecommand \@@endlink[0]{}%
\providecommand \url  [0]{\begingroup\@sanitize@url \@url }%
\providecommand \@url [1]{\endgroup\@href {#1}{\urlprefix }}%
\providecommand \urlprefix  [0]{URL }%
\providecommand \Eprint [0]{\href }%
\providecommand \doibase [0]{http://dx.doi.org/}%
\providecommand \selectlanguage [0]{\@gobble}%
\providecommand \bibinfo  [0]{\@secondoftwo}%
\providecommand \bibfield  [0]{\@secondoftwo}%
\providecommand \translation [1]{[#1]}%
\providecommand \BibitemOpen [0]{}%
\providecommand \bibitemStop [0]{}%
\providecommand \bibitemNoStop [0]{.\EOS\space}%
\providecommand \EOS [0]{\spacefactor3000\relax}%
\providecommand \BibitemShut  [1]{\csname bibitem#1\endcsname}%
\let\auto@bib@innerbib\@empty
\bibitem [{\citenamefont {Hikami}\ \emph {et~al.}(1980)\citenamefont {Hikami},
  \citenamefont {Larkin},\ and\ \citenamefont {Nagaoka}}]{Hikami1980}%
  \BibitemOpen
  \bibfield  {author} {\bibinfo {author} {\bibfnamefont {S.}~\bibnamefont
  {Hikami}}, \bibinfo {author} {\bibfnamefont {A.~I.}\ \bibnamefont {Larkin}},
  \ and\ \bibinfo {author} {\bibfnamefont {Y.}~\bibnamefont {Nagaoka}},\ }\href
  {\doibase 10.1143/PTP.63.707} {\bibfield  {journal} {\bibinfo  {journal}
  {Prog. Theor. Phys.}\ }\textbf {\bibinfo {volume} {63}},\ \bibinfo {pages}
  {707} (\bibinfo {year} {1980})}\BibitemShut {NoStop}%
\bibitem [{\citenamefont {Gorkov}\ \emph {et~al.}(1979)\citenamefont {Gorkov},
  \citenamefont {Larkin},\ and\ \citenamefont {Khmel’nitskii}}]{Gorkov1979}%
  \BibitemOpen
  \bibfield  {author} {\bibinfo {author} {\bibfnamefont {L.~P.}\ \bibnamefont
  {Gorkov}}, \bibinfo {author} {\bibfnamefont {A.~I.}\ \bibnamefont {Larkin}},
  \ and\ \bibinfo {author} {\bibfnamefont {D.~E.}\ \bibnamefont
  {Khmel’nitskii}},\ }\href@noop {} {\bibfield  {journal} {\bibinfo
  {journal} {Pis’ma Zh. Eksp. Teor. Fiz.}\ }\textbf {\bibinfo {volume}
  {30}},\ \bibinfo {pages} {248} (\bibinfo {year} {1979})},\ \bibinfo {note}
  {[JETP Lett. 30, 228 (1979)]}\BibitemShut {NoStop}%
\bibitem [{\citenamefont {Anderson}\ \emph {et~al.}(1979)\citenamefont
  {Anderson}, \citenamefont {Abrahams},\ and\ \citenamefont
  {Ramakrishnan}}]{Anderson1979}%
  \BibitemOpen
  \bibfield  {author} {\bibinfo {author} {\bibfnamefont {P.~W.}\ \bibnamefont
  {Anderson}}, \bibinfo {author} {\bibfnamefont {E.}~\bibnamefont {Abrahams}},
  \ and\ \bibinfo {author} {\bibfnamefont {T.~V.}\ \bibnamefont
  {Ramakrishnan}},\ }\href {\doibase 10.1103/PhysRevLett.43.718} {\bibfield
  {journal} {\bibinfo  {journal} {Phys. Rev. Lett.}\ }\textbf {\bibinfo
  {volume} {43}},\ \bibinfo {pages} {718} (\bibinfo {year} {1979})}\BibitemShut
  {NoStop}%
\bibitem [{\citenamefont {Knap}\ \emph {et~al.}(1996)\citenamefont {Knap},
  \citenamefont {Skierbiszewski}, \citenamefont {Zduniak}, \citenamefont
  {Litwin-Staszewska}, \citenamefont {Bertho}, \citenamefont {Kobbi},
  \citenamefont {Robert}, \citenamefont {Pikus}, \citenamefont {Pikus},
  \citenamefont {Iordanskii}, \citenamefont {Mosser}, \citenamefont
  {Zekentes},\ and\ \citenamefont {\mbox{Yu}. B.~Lyanda-Geller}}]{Knap1996}%
  \BibitemOpen
  \bibfield  {author} {\bibinfo {author} {\bibfnamefont {W.}~\bibnamefont
  {Knap}}, \bibinfo {author} {\bibfnamefont {C.}~\bibnamefont
  {Skierbiszewski}}, \bibinfo {author} {\bibfnamefont {A.}~\bibnamefont
  {Zduniak}}, \bibinfo {author} {\bibfnamefont {E.}~\bibnamefont
  {Litwin-Staszewska}}, \bibinfo {author} {\bibfnamefont {D.}~\bibnamefont
  {Bertho}}, \bibinfo {author} {\bibfnamefont {F.}~\bibnamefont {Kobbi}},
  \bibinfo {author} {\bibfnamefont {J.~L.}\ \bibnamefont {Robert}}, \bibinfo
  {author} {\bibfnamefont {G.~E.}\ \bibnamefont {Pikus}}, \bibinfo {author}
  {\bibfnamefont {F.~G.}\ \bibnamefont {Pikus}}, \bibinfo {author}
  {\bibfnamefont {S.~V.}\ \bibnamefont {Iordanskii}}, \bibinfo {author}
  {\bibfnamefont {V.}~\bibnamefont {Mosser}}, \bibinfo {author} {\bibfnamefont
  {K.}~\bibnamefont {Zekentes}}, \ and\ \bibinfo {author} {\bibnamefont
  {\mbox{Yu}. B.~Lyanda-Geller}},\ }\href
  {http://link.aps.org/doi/10.1103/PhysRevB.53.3912} {\bibfield  {journal}
  {\bibinfo  {journal} {Phys. Rev. B}\ }\textbf {\bibinfo {volume} {53}},\
  \bibinfo {pages} {3912} (\bibinfo {year} {1996})}\BibitemShut {NoStop}%
\bibitem [{\citenamefont {Hassenkam}\ \emph {et~al.}(1997)\citenamefont
  {Hassenkam}, \citenamefont {Pedersen}, \citenamefont {Baklanov},
  \citenamefont {Kristensen}, \citenamefont {Sorensen}, \citenamefont
  {Lindelof}, \citenamefont {Pikus},\ and\ \citenamefont
  {Pikus}}]{Hassenkam1997}%
  \BibitemOpen
  \bibfield  {author} {\bibinfo {author} {\bibfnamefont {T.}~\bibnamefont
  {Hassenkam}}, \bibinfo {author} {\bibfnamefont {S.}~\bibnamefont {Pedersen}},
  \bibinfo {author} {\bibfnamefont {K.}~\bibnamefont {Baklanov}}, \bibinfo
  {author} {\bibfnamefont {A.}~\bibnamefont {Kristensen}}, \bibinfo {author}
  {\bibfnamefont {C.~B.}\ \bibnamefont {Sorensen}}, \bibinfo {author}
  {\bibfnamefont {P.~E.}\ \bibnamefont {Lindelof}}, \bibinfo {author}
  {\bibfnamefont {F.~G.}\ \bibnamefont {Pikus}}, \ and\ \bibinfo {author}
  {\bibfnamefont {G.~E.}\ \bibnamefont {Pikus}},\ }\href {\doibase
  10.1103/PhysRevB.55.9298} {\bibfield  {journal} {\bibinfo  {journal} {Phys.
  Rev. B}\ }\textbf {\bibinfo {volume} {55}},\ \bibinfo {pages} {9298}
  (\bibinfo {year} {1997})}\BibitemShut {NoStop}%
\bibitem [{\citenamefont {Zumb\"uhl}\ \emph {et~al.}(2002)\citenamefont
  {Zumb\"uhl}, \citenamefont {Miller}, \citenamefont {Marcus}, \citenamefont
  {Campman},\ and\ \citenamefont {Gossard}}]{Zumbuehl2002}%
  \BibitemOpen
  \bibfield  {author} {\bibinfo {author} {\bibfnamefont {D.~M.}\ \bibnamefont
  {Zumb\"uhl}}, \bibinfo {author} {\bibfnamefont {J.~B.}\ \bibnamefont
  {Miller}}, \bibinfo {author} {\bibfnamefont {C.~M.}\ \bibnamefont {Marcus}},
  \bibinfo {author} {\bibfnamefont {K.}~\bibnamefont {Campman}}, \ and\
  \bibinfo {author} {\bibfnamefont {A.~C.}\ \bibnamefont {Gossard}},\ }\href
  {\doibase 10.1103/PhysRevLett.89.276803} {\bibfield  {journal} {\bibinfo
  {journal} {Phys. Rev. Lett.}\ }\textbf {\bibinfo {volume} {89}},\ \bibinfo
  {pages} {276803} (\bibinfo {year} {2002})}\BibitemShut {NoStop}%
\bibitem [{\citenamefont {\mbox{Th}. Sch\"apers}\ \emph
  {et~al.}(2006)\citenamefont {\mbox{Th}. Sch\"apers}, \citenamefont {Guzenko},
  \citenamefont {Pala}, \citenamefont {Z\"ulicke}, \citenamefont {Governale},
  \citenamefont {Knobbe},\ and\ \citenamefont {Hardtdegen}}]{Schapers2006}%
  \BibitemOpen
  \bibfield  {author} {\bibinfo {author} {\bibnamefont {\mbox{Th}.
  Sch\"apers}}, \bibinfo {author} {\bibfnamefont {V.~A.}\ \bibnamefont
  {Guzenko}}, \bibinfo {author} {\bibfnamefont {M.~G.}\ \bibnamefont {Pala}},
  \bibinfo {author} {\bibfnamefont {U.}~\bibnamefont {Z\"ulicke}}, \bibinfo
  {author} {\bibfnamefont {M.}~\bibnamefont {Governale}}, \bibinfo {author}
  {\bibfnamefont {J.}~\bibnamefont {Knobbe}}, \ and\ \bibinfo {author}
  {\bibfnamefont {H.}~\bibnamefont {Hardtdegen}},\ }\href {\doibase
  10.1103/PhysRevB.74.081301} {\bibfield  {journal} {\bibinfo  {journal} {Phys.
  Rev. B}\ }\textbf {\bibinfo {volume} {74}},\ \bibinfo {pages} {081301(R)}
  (\bibinfo {year} {2006})}\BibitemShut {NoStop}%
\bibitem [{\citenamefont {Thillosen}\ \emph {et~al.}(2006)\citenamefont
  {Thillosen}, \citenamefont {Caba\~nas}, \citenamefont {Kaluza}, \citenamefont
  {Guzenko}, \citenamefont {Hardtdegen},\ and\ \citenamefont {\mbox{Th}.
  Sch\"apers}}]{Thillosen2006}%
  \BibitemOpen
  \bibfield  {author} {\bibinfo {author} {\bibfnamefont {N.}~\bibnamefont
  {Thillosen}}, \bibinfo {author} {\bibfnamefont {S.}~\bibnamefont
  {Caba\~nas}}, \bibinfo {author} {\bibfnamefont {N.}~\bibnamefont {Kaluza}},
  \bibinfo {author} {\bibfnamefont {V.~A.}\ \bibnamefont {Guzenko}}, \bibinfo
  {author} {\bibfnamefont {H.}~\bibnamefont {Hardtdegen}}, \ and\ \bibinfo
  {author} {\bibnamefont {\mbox{Th}. Sch\"apers}},\ }\href {\doibase
  10.1103/PhysRevB.73.241311} {\bibfield  {journal} {\bibinfo  {journal} {Phys.
  Rev. B}\ }\textbf {\bibinfo {volume} {73}},\ \bibinfo {pages} {241311(R)}
  (\bibinfo {year} {2006})}\BibitemShut {NoStop}%
\bibitem [{\citenamefont {Guzenko}\ \emph {et~al.}(2007)\citenamefont
  {Guzenko}, \citenamefont {\mbox{Th.} Sch\"apers},\ and\ \citenamefont
  {Hardtdegen}}]{Guzenko2007}%
  \BibitemOpen
  \bibfield  {author} {\bibinfo {author} {\bibfnamefont {V.~A.}\ \bibnamefont
  {Guzenko}}, \bibinfo {author} {\bibnamefont {\mbox{Th.} Sch\"apers}}, \ and\
  \bibinfo {author} {\bibfnamefont {H.}~\bibnamefont {Hardtdegen}},\ }\href
  {\doibase 10.1103/PhysRevB.76.165301} {\bibfield  {journal} {\bibinfo
  {journal} {Phys. Rev. B}\ }\textbf {\bibinfo {volume} {76}},\ \bibinfo
  {pages} {165301} (\bibinfo {year} {2007})}\BibitemShut {NoStop}%
\bibitem [{\citenamefont {Kunihashi}\ \emph {et~al.}(2009)\citenamefont
  {Kunihashi}, \citenamefont {Kohda},\ and\ \citenamefont
  {Nitta}}]{Kunihashi2009}%
  \BibitemOpen
  \bibfield  {author} {\bibinfo {author} {\bibfnamefont {Y.}~\bibnamefont
  {Kunihashi}}, \bibinfo {author} {\bibfnamefont {M.}~\bibnamefont {Kohda}}, \
  and\ \bibinfo {author} {\bibfnamefont {J.}~\bibnamefont {Nitta}},\ }\href
  {\doibase 10.1103/PhysRevLett.102.226601} {\bibfield  {journal} {\bibinfo
  {journal} {Phys. Rev. Lett.}\ }\textbf {\bibinfo {volume} {102}},\ \bibinfo
  {pages} {226601 } (\bibinfo {year} {2009})}\BibitemShut {NoStop}%
\bibitem [{\citenamefont {Caviglia}\ \emph {et~al.}(2010)\citenamefont
  {Caviglia}, \citenamefont {Gabay}, \citenamefont {Gariglio}, \citenamefont
  {Reyren}, \citenamefont {Cancellieri},\ and\ \citenamefont
  {Triscone}}]{Caviglia2010}%
  \BibitemOpen
  \bibfield  {author} {\bibinfo {author} {\bibfnamefont {A.~D.}\ \bibnamefont
  {Caviglia}}, \bibinfo {author} {\bibfnamefont {M.}~\bibnamefont {Gabay}},
  \bibinfo {author} {\bibfnamefont {S.}~\bibnamefont {Gariglio}}, \bibinfo
  {author} {\bibfnamefont {N.}~\bibnamefont {Reyren}}, \bibinfo {author}
  {\bibfnamefont {C.}~\bibnamefont {Cancellieri}}, \ and\ \bibinfo {author}
  {\bibfnamefont {J.-M.}\ \bibnamefont {Triscone}},\ }\href {\doibase
  10.1103/PhysRevLett.104.126803} {\bibfield  {journal} {\bibinfo  {journal}
  {Phys. Rev. Lett.}\ }\textbf {\bibinfo {volume} {104}},\ \bibinfo {pages}
  {126803} (\bibinfo {year} {2010})}\BibitemShut {NoStop}%
\bibitem [{\citenamefont {Moriya}\ \emph {et~al.}(2014)\citenamefont {Moriya},
  \citenamefont {Sawano}, \citenamefont {Hoshi}, \citenamefont {Masubuchi},
  \citenamefont {Shiraki}, \citenamefont {Wild}, \citenamefont {Neumann},
  \citenamefont {Abstreiter}, \citenamefont {Bougeard}, \citenamefont {Koga},\
  and\ \citenamefont {Machida}}]{Moriya2014}%
  \BibitemOpen
  \bibfield  {author} {\bibinfo {author} {\bibfnamefont {R.}~\bibnamefont
  {Moriya}}, \bibinfo {author} {\bibfnamefont {K.}~\bibnamefont {Sawano}},
  \bibinfo {author} {\bibfnamefont {Y.}~\bibnamefont {Hoshi}}, \bibinfo
  {author} {\bibfnamefont {S.}~\bibnamefont {Masubuchi}}, \bibinfo {author}
  {\bibfnamefont {Y.}~\bibnamefont {Shiraki}}, \bibinfo {author} {\bibfnamefont
  {A.}~\bibnamefont {Wild}}, \bibinfo {author} {\bibfnamefont {C.}~\bibnamefont
  {Neumann}}, \bibinfo {author} {\bibfnamefont {G.}~\bibnamefont {Abstreiter}},
  \bibinfo {author} {\bibfnamefont {D.}~\bibnamefont {Bougeard}}, \bibinfo
  {author} {\bibfnamefont {T.}~\bibnamefont {Koga}}, \ and\ \bibinfo {author}
  {\bibfnamefont {T.}~\bibnamefont {Machida}},\ }\href {\doibase
  10.1103/PhysRevLett.113.086601} {\bibfield  {journal} {\bibinfo  {journal}
  {Phys. Rev. Lett.}\ }\textbf {\bibinfo {volume} {113}},\ \bibinfo {pages}
  {086601} (\bibinfo {year} {2014})}\BibitemShut {NoStop}%
\bibitem [{\citenamefont {van Weperen}\ \emph {et~al.}(2015)\citenamefont {van
  Weperen}, \citenamefont {Tarasinski}, \citenamefont {Eeltink}, \citenamefont
  {Pribiag}, \citenamefont {Plissard}, \citenamefont {Bakkers}, \citenamefont
  {Kouwenhoven},\ and\ \citenamefont {Wimmer}}]{Weperen2015}%
  \BibitemOpen
  \bibfield  {author} {\bibinfo {author} {\bibfnamefont {I.}~\bibnamefont {van
  Weperen}}, \bibinfo {author} {\bibfnamefont {B.}~\bibnamefont {Tarasinski}},
  \bibinfo {author} {\bibfnamefont {D.}~\bibnamefont {Eeltink}}, \bibinfo
  {author} {\bibfnamefont {V.~S.}\ \bibnamefont {Pribiag}}, \bibinfo {author}
  {\bibfnamefont {S.~R.}\ \bibnamefont {Plissard}}, \bibinfo {author}
  {\bibfnamefont {E.~P. A.~M.}\ \bibnamefont {Bakkers}}, \bibinfo {author}
  {\bibfnamefont {L.~P.}\ \bibnamefont {Kouwenhoven}}, \ and\ \bibinfo {author}
  {\bibfnamefont {M.}~\bibnamefont {Wimmer}},\ }\href {\doibase
  10.1103/PhysRevB.91.201413} {\bibfield  {journal} {\bibinfo  {journal} {Phys.
  Rev. B}\ }\textbf {\bibinfo {volume} {91}},\ \bibinfo {pages} {201413(R)}
  (\bibinfo {year} {2015})}\BibitemShut {NoStop}%
\bibitem [{\citenamefont {Kammermeier}\ \emph
  {et~al.}(2016{\natexlab{a}})\citenamefont {Kammermeier}, \citenamefont
  {Wenk}, \citenamefont {Schliemann}, \citenamefont {Heedt},\ and\
  \citenamefont {\mbox{Th}. Sch\"apers}}]{Kammermeier2016}%
  \BibitemOpen
  \bibfield  {author} {\bibinfo {author} {\bibfnamefont {M.}~\bibnamefont
  {Kammermeier}}, \bibinfo {author} {\bibfnamefont {P.}~\bibnamefont {Wenk}},
  \bibinfo {author} {\bibfnamefont {J.}~\bibnamefont {Schliemann}}, \bibinfo
  {author} {\bibfnamefont {S.}~\bibnamefont {Heedt}}, \ and\ \bibinfo {author}
  {\bibnamefont {\mbox{Th}. Sch\"apers}},\ }\href {\doibase
  10.1103/PhysRevB.93.205306} {\bibfield  {journal} {\bibinfo  {journal} {Phys.
  Rev. B}\ }\textbf {\bibinfo {volume} {93}},\ \bibinfo {pages} {205306}
  (\bibinfo {year} {2016}{\natexlab{a}})}\BibitemShut {NoStop}%
\bibitem [{\citenamefont {Kammermeier}\ \emph {et~al.}(2017)\citenamefont
  {Kammermeier}, \citenamefont {Wenk}, \citenamefont {Schliemann},
  \citenamefont {Heedt}, \citenamefont {\mbox{Th}. Gerster},\ and\
  \citenamefont {\mbox{Th}. Sch\"apers}}]{Kammermeier2017}%
  \BibitemOpen
  \bibfield  {author} {\bibinfo {author} {\bibfnamefont {M.}~\bibnamefont
  {Kammermeier}}, \bibinfo {author} {\bibfnamefont {P.}~\bibnamefont {Wenk}},
  \bibinfo {author} {\bibfnamefont {J.}~\bibnamefont {Schliemann}}, \bibinfo
  {author} {\bibfnamefont {S.}~\bibnamefont {Heedt}}, \bibinfo {author}
  {\bibnamefont {\mbox{Th}. Gerster}}, \ and\ \bibinfo {author} {\bibnamefont
  {\mbox{Th}. Sch\"apers}},\ }\href {\doibase 10.1103/PhysRevB.96.235302}
  {\bibfield  {journal} {\bibinfo  {journal} {Phys. Rev. B}\ }\textbf {\bibinfo
  {volume} {96}},\ \bibinfo {pages} {235302} (\bibinfo {year}
  {2017})}\BibitemShut {NoStop}%
\bibitem [{\citenamefont {Jespersen}\ \emph {et~al.}(2018)\citenamefont
  {Jespersen}, \citenamefont {Krogstrup}, \citenamefont {Lunde}, \citenamefont
  {Tanta}, \citenamefont {Kanne}, \citenamefont {Johnson},\ and\ \citenamefont
  {Nyg\aa{}rd}}]{Jespersen2018}%
  \BibitemOpen
  \bibfield  {author} {\bibinfo {author} {\bibfnamefont {T.~S.}\ \bibnamefont
  {Jespersen}}, \bibinfo {author} {\bibfnamefont {P.}~\bibnamefont
  {Krogstrup}}, \bibinfo {author} {\bibfnamefont {A.~M.}\ \bibnamefont
  {Lunde}}, \bibinfo {author} {\bibfnamefont {R.}~\bibnamefont {Tanta}},
  \bibinfo {author} {\bibfnamefont {T.}~\bibnamefont {Kanne}}, \bibinfo
  {author} {\bibfnamefont {E.}~\bibnamefont {Johnson}}, \ and\ \bibinfo
  {author} {\bibfnamefont {J.}~\bibnamefont {Nyg\aa{}rd}},\ }\href {\doibase
  10.1103/PhysRevB.97.041303} {\bibfield  {journal} {\bibinfo  {journal} {Phys.
  Rev. B}\ }\textbf {\bibinfo {volume} {97}},\ \bibinfo {pages} {041303(R)}
  (\bibinfo {year} {2018})}\BibitemShut {NoStop}%
\bibitem [{\citenamefont {Elliott}(1954)}]{elliott}%
  \BibitemOpen
  \bibfield  {author} {\bibinfo {author} {\bibfnamefont {R.~J.}\ \bibnamefont
  {Elliott}},\ }\href {\doibase 10.1103/PhysRev.96.266} {\bibfield  {journal}
  {\bibinfo  {journal} {Phys. Rev.}\ }\textbf {\bibinfo {volume} {96}},\
  \bibinfo {pages} {266} (\bibinfo {year} {1954})}\BibitemShut {NoStop}%
\bibitem [{\citenamefont {Yafet}(1963)}]{yafet}%
  \BibitemOpen
  \bibfield  {author} {\bibinfo {author} {\bibfnamefont {Y.}~\bibnamefont
  {Yafet}},\ }in\ \href@noop {} {\emph {\bibinfo {booktitle} {Solid State
  Physics}}},\ Vol.~\bibinfo {volume} {14},\ \bibinfo {editor} {edited by\
  \bibinfo {editor} {\bibfnamefont {F.}~\bibnamefont {Seitz}}\ and\ \bibinfo
  {editor} {\bibfnamefont {D.}~\bibnamefont {Turnbull}}}\ (\bibinfo
  {publisher} {Academic, New York},\ \bibinfo {year} {1963})\BibitemShut
  {NoStop}%
\bibitem [{\citenamefont {D'yakonov}\ and\ \citenamefont
  {Perel'}(1972)}]{perel}%
  \BibitemOpen
  \bibfield  {author} {\bibinfo {author} {\bibfnamefont {M.~I.}\ \bibnamefont
  {D'yakonov}}\ and\ \bibinfo {author} {\bibfnamefont {V.~I.}\ \bibnamefont
  {Perel'}},\ }\href@noop {} {\bibfield  {journal} {\bibinfo  {journal} {Sov.
  Phys. Solid State}\ }\textbf {\bibinfo {volume} {13}},\ \bibinfo {pages}
  {3023} (\bibinfo {year} {1972})},\ \bibinfo {note} {[Fiz. Tverd. Tela {\bf
  13}, 3581 (1971)]}\BibitemShut {NoStop}%
\bibitem [{\citenamefont {Rashba}(1960)}]{Rashba1960}%
  \BibitemOpen
  \bibfield  {author} {\bibinfo {author} {\bibfnamefont {E.~I.}\ \bibnamefont
  {Rashba}},\ }\href@noop {} {\bibfield  {journal} {\bibinfo  {journal} {Sov.
  Phys. Solid State}\ }\textbf {\bibinfo {volume} {2}},\ \bibinfo {pages}
  {1109} (\bibinfo {year} {1960})},\ \bibinfo {note} {[Fiz. Tverd. Tela {\bf
  2}, 1224 (1960)]}\BibitemShut {NoStop}%
\bibitem [{\citenamefont {Bychkov}\ and\ \citenamefont
  {Rashba}(1984)}]{Bychkov1984}%
  \BibitemOpen
  \bibfield  {author} {\bibinfo {author} {\bibfnamefont {Y.~A.}\ \bibnamefont
  {Bychkov}}\ and\ \bibinfo {author} {\bibfnamefont {E.~I.}\ \bibnamefont
  {Rashba}},\ }\href {\doibase 10.1088/0022-3719/17/33/015} {\bibfield
  {journal} {\bibinfo  {journal} {J. Phys. C: Solid State Phys.}\ }\textbf
  {\bibinfo {volume} {17}},\ \bibinfo {pages} {6039} (\bibinfo {year}
  {1984})}\BibitemShut {NoStop}%
\bibitem [{\citenamefont {Dresselhaus}(1955)}]{Dresselhaus1955}%
  \BibitemOpen
  \bibfield  {author} {\bibinfo {author} {\bibfnamefont {G.}~\bibnamefont
  {Dresselhaus}},\ }\href {\doibase 10.1103/PhysRev.100.580} {\bibfield
  {journal} {\bibinfo  {journal} {Phys. Rev.}\ }\textbf {\bibinfo {volume}
  {100}},\ \bibinfo {pages} {580} (\bibinfo {year} {1955})}\BibitemShut
  {NoStop}%
\bibitem [{\citenamefont {Pikus}\ and\ \citenamefont
  {Pikus}(1995)}]{Pikus1995}%
  \BibitemOpen
  \bibfield  {author} {\bibinfo {author} {\bibfnamefont {F.~G.}\ \bibnamefont
  {Pikus}}\ and\ \bibinfo {author} {\bibfnamefont {G.~E.}\ \bibnamefont
  {Pikus}},\ }\href {http://link.aps.org/doi/10.1103/PhysRevB.51.16928}
  {\bibfield  {journal} {\bibinfo  {journal} {Phys. Rev. B}\ }\textbf {\bibinfo
  {volume} {51}},\ \bibinfo {pages} {16928} (\bibinfo {year}
  {1995})}\BibitemShut {NoStop}%
\bibitem [{\citenamefont {Altshuler}\ \emph {et~al.}(1981)\citenamefont
  {Altshuler}, \citenamefont {Aronov}, \citenamefont {Larkin},\ and\
  \citenamefont {Khmelnitskii}}]{Altshuler1981b}%
  \BibitemOpen
  \bibfield  {author} {\bibinfo {author} {\bibfnamefont {B.~L.}\ \bibnamefont
  {Altshuler}}, \bibinfo {author} {\bibfnamefont {A.~G.}\ \bibnamefont
  {Aronov}}, \bibinfo {author} {\bibfnamefont {A.~I.}\ \bibnamefont {Larkin}},
  \ and\ \bibinfo {author} {\bibfnamefont {D.~E.}\ \bibnamefont
  {Khmelnitskii}},\ }\href@noop {} {\bibfield  {journal} {\bibinfo  {journal}
  {Sov. Phys. JETP}\ }\textbf {\bibinfo {volume} {54}},\ \bibinfo {pages} {411}
  (\bibinfo {year} {1981})},\ \bibinfo {note} {[Zh. Eksp. Teor. Fiz. {\bf 81},
  768 (1981)]}\BibitemShut {NoStop}%
\bibitem [{\citenamefont {Iordanskii}\ \emph {et~al.}(1994)\citenamefont
  {Iordanskii}, \citenamefont {\mbox{Yu}. B.~Lyanda-Geller},\ and\
  \citenamefont {Pikus}}]{Iordanskii1994}%
  \BibitemOpen
  \bibfield  {author} {\bibinfo {author} {\bibfnamefont {S.~V.}\ \bibnamefont
  {Iordanskii}}, \bibinfo {author} {\bibnamefont {\mbox{Yu}.
  B.~Lyanda-Geller}}, \ and\ \bibinfo {author} {\bibfnamefont {G.~E.}\
  \bibnamefont {Pikus}},\ }\href
  {http://www.jetpletters.ac.ru/ps/1323/article\_20010.pdf} {\bibfield
  {journal} {\bibinfo  {journal} {JETP Lett.}\ }\textbf {\bibinfo {volume}
  {60}},\ \bibinfo {pages} {206} (\bibinfo {year} {1994})},\ \bibinfo {note}
  {[Pis'ma Zh. Eksp. Teor. Fiz. {\bf 60}, 199 (1994)]}\BibitemShut {NoStop}%
\bibitem [{\citenamefont {Schliemann}\ \emph {et~al.}(2003)\citenamefont
  {Schliemann}, \citenamefont {Egues},\ and\ \citenamefont
  {Loss}}]{Schliemann2003}%
  \BibitemOpen
  \bibfield  {author} {\bibinfo {author} {\bibfnamefont {J.}~\bibnamefont
  {Schliemann}}, \bibinfo {author} {\bibfnamefont {J.~C.}\ \bibnamefont
  {Egues}}, \ and\ \bibinfo {author} {\bibfnamefont {D.}~\bibnamefont {Loss}},\
  }\href {\doibase 10.1103/PhysRevLett.90.146801} {\bibfield  {journal}
  {\bibinfo  {journal} {Phys. Rev. Lett.}\ }\textbf {\bibinfo {volume} {90}},\
  \bibinfo {pages} {146801} (\bibinfo {year} {2003})}\BibitemShut {NoStop}%
\bibitem [{\citenamefont {Bernevig}\ \emph {et~al.}(2006)\citenamefont
  {Bernevig}, \citenamefont {Orenstein},\ and\ \citenamefont
  {Zhang}}]{Bernevig2006}%
  \BibitemOpen
  \bibfield  {author} {\bibinfo {author} {\bibfnamefont {B.~A.}\ \bibnamefont
  {Bernevig}}, \bibinfo {author} {\bibfnamefont {J.}~\bibnamefont {Orenstein}},
  \ and\ \bibinfo {author} {\bibfnamefont {S.-C.}\ \bibnamefont {Zhang}},\
  }\href {\doibase 10.1103/PhysRevLett.97.236601} {\bibfield  {journal}
  {\bibinfo  {journal} {Phys. Rev. Lett.}\ }\textbf {\bibinfo {volume} {97}},\
  \bibinfo {pages} {236601} (\bibinfo {year} {2006})}\BibitemShut {NoStop}%
\bibitem [{\citenamefont {Schliemann}(2017)}]{Schliemann2017}%
  \BibitemOpen
  \bibfield  {author} {\bibinfo {author} {\bibfnamefont {J.}~\bibnamefont
  {Schliemann}},\ }\href {\doibase 10.1103/RevModPhys.89.011001} {\bibfield
  {journal} {\bibinfo  {journal} {Rev. Mod. Phys.}\ }\textbf {\bibinfo {volume}
  {89}},\ \bibinfo {pages} {011001} (\bibinfo {year} {2017})}\BibitemShut
  {NoStop}%
\bibitem [{\citenamefont {Kohda}\ and\ \citenamefont
  {Salis}(2017)}]{Kohda2017}%
  \BibitemOpen
  \bibfield  {author} {\bibinfo {author} {\bibfnamefont {M.}~\bibnamefont
  {Kohda}}\ and\ \bibinfo {author} {\bibfnamefont {G.}~\bibnamefont {Salis}},\
  }\href {http://stacks.iop.org/0268-1242/32/i=7/a=073002} {\bibfield
  {journal} {\bibinfo  {journal} {Semicond. Sci. Technol.}\ }\textbf {\bibinfo
  {volume} {32}},\ \bibinfo {pages} {073002} (\bibinfo {year}
  {2017})}\BibitemShut {NoStop}%
\bibitem [{\citenamefont {Kammermeier}\ \emph
  {et~al.}(2016{\natexlab{b}})\citenamefont {Kammermeier}, \citenamefont
  {Wenk},\ and\ \citenamefont {Schliemann}}]{Kammermeier2016PRL}%
  \BibitemOpen
  \bibfield  {author} {\bibinfo {author} {\bibfnamefont {M.}~\bibnamefont
  {Kammermeier}}, \bibinfo {author} {\bibfnamefont {P.}~\bibnamefont {Wenk}}, \
  and\ \bibinfo {author} {\bibfnamefont {J.}~\bibnamefont {Schliemann}},\
  }\href {\doibase 10.1103/PhysRevLett.117.236801} {\bibfield  {journal}
  {\bibinfo  {journal} {Phys. Rev. Lett.}\ }\textbf {\bibinfo {volume} {117}},\
  \bibinfo {pages} {236801} (\bibinfo {year} {2016}{\natexlab{b}})}\BibitemShut
  {NoStop}%
\bibitem [{\citenamefont {Marinescu}\ \emph {et~al.}(2019)\citenamefont
  {Marinescu}, \citenamefont {Weigele}, \citenamefont {Zumb\"uhl},\ and\
  \citenamefont {Egues}}]{Marinescu2019}%
  \BibitemOpen
  \bibfield  {author} {\bibinfo {author} {\bibfnamefont {D.~C.}\ \bibnamefont
  {Marinescu}}, \bibinfo {author} {\bibfnamefont {P.~J.}\ \bibnamefont
  {Weigele}}, \bibinfo {author} {\bibfnamefont {D.~M.}\ \bibnamefont
  {Zumb\"uhl}}, \ and\ \bibinfo {author} {\bibfnamefont {J.~C.}\ \bibnamefont
  {Egues}},\ }\href {\doibase 10.1103/PhysRevLett.122.156601} {\bibfield
  {journal} {\bibinfo  {journal} {Phys. Rev. Lett.}\ }\textbf {\bibinfo
  {volume} {122}},\ \bibinfo {pages} {156601} (\bibinfo {year}
  {2019})}\BibitemShut {NoStop}%
\bibitem [{\citenamefont {Weigele}\ \emph {et~al.}(2020)\citenamefont
  {Weigele}, \citenamefont {Marinescu}, \citenamefont {Dettwiler},
  \citenamefont {Fu}, \citenamefont {Mack}, \citenamefont {Egues},
  \citenamefont {Awschalom},\ and\ \citenamefont {Zumb\"uhl}}]{Weigele2020}%
  \BibitemOpen
  \bibfield  {author} {\bibinfo {author} {\bibfnamefont {P.~J.}\ \bibnamefont
  {Weigele}}, \bibinfo {author} {\bibfnamefont {D.~C.}\ \bibnamefont
  {Marinescu}}, \bibinfo {author} {\bibfnamefont {F.}~\bibnamefont
  {Dettwiler}}, \bibinfo {author} {\bibfnamefont {J.}~\bibnamefont {Fu}},
  \bibinfo {author} {\bibfnamefont {S.}~\bibnamefont {Mack}}, \bibinfo {author}
  {\bibfnamefont {J.~C.}\ \bibnamefont {Egues}}, \bibinfo {author}
  {\bibfnamefont {D.~D.}\ \bibnamefont {Awschalom}}, \ and\ \bibinfo {author}
  {\bibfnamefont {D.~M.}\ \bibnamefont {Zumb\"uhl}},\ }\href {\doibase
  10.1103/PhysRevB.101.035414} {\bibfield  {journal} {\bibinfo  {journal}
  {Phys. Rev. B}\ }\textbf {\bibinfo {volume} {101}},\ \bibinfo {pages}
  {035414} (\bibinfo {year} {2020})}\BibitemShut {NoStop}%
\bibitem [{\citenamefont {Kohda}\ \emph {et~al.}(2012)\citenamefont {Kohda},
  \citenamefont {Lechner}, \citenamefont {Kunihashi}, \citenamefont
  {Dollinger}, \citenamefont {Olbrich}, \citenamefont {Sch\"onhuber},
  \citenamefont {Caspers}, \citenamefont {Bel'kov}, \citenamefont {Golub},
  \citenamefont {Weiss}, \citenamefont {Richter}, \citenamefont {Nitta},\ and\
  \citenamefont {Ganichev}}]{Kohda2012}%
  \BibitemOpen
  \bibfield  {author} {\bibinfo {author} {\bibfnamefont {M.}~\bibnamefont
  {Kohda}}, \bibinfo {author} {\bibfnamefont {V.}~\bibnamefont {Lechner}},
  \bibinfo {author} {\bibfnamefont {Y.}~\bibnamefont {Kunihashi}}, \bibinfo
  {author} {\bibfnamefont {T.}~\bibnamefont {Dollinger}}, \bibinfo {author}
  {\bibfnamefont {P.}~\bibnamefont {Olbrich}}, \bibinfo {author} {\bibfnamefont
  {C.}~\bibnamefont {Sch\"onhuber}}, \bibinfo {author} {\bibfnamefont
  {I.}~\bibnamefont {Caspers}}, \bibinfo {author} {\bibfnamefont {V.~V.}\
  \bibnamefont {Bel'kov}}, \bibinfo {author} {\bibfnamefont {L.~E.}\
  \bibnamefont {Golub}}, \bibinfo {author} {\bibfnamefont {D.}~\bibnamefont
  {Weiss}}, \bibinfo {author} {\bibfnamefont {K.}~\bibnamefont {Richter}},
  \bibinfo {author} {\bibfnamefont {J.}~\bibnamefont {Nitta}}, \ and\ \bibinfo
  {author} {\bibfnamefont {S.~D.}\ \bibnamefont {Ganichev}},\ }\href {\doibase
  10.1103/PhysRevB.86.081306} {\bibfield  {journal} {\bibinfo  {journal} {Phys.
  Rev. B}\ }\textbf {\bibinfo {volume} {86}},\ \bibinfo {pages} {081306(R)}
  (\bibinfo {year} {2012})}\BibitemShut {NoStop}%
\bibitem [{\citenamefont {Iizasa}\ \emph {et~al.}(2020)\citenamefont {Iizasa},
  \citenamefont {Kohda}, \citenamefont {Z\"ulicke}, \citenamefont {Nitta},\
  and\ \citenamefont {Kammermeier}}]{Iizasa2020}%
  \BibitemOpen
  \bibfield  {author} {\bibinfo {author} {\bibfnamefont {D.}~\bibnamefont
  {Iizasa}}, \bibinfo {author} {\bibfnamefont {M.}~\bibnamefont {Kohda}},
  \bibinfo {author} {\bibfnamefont {U.}~\bibnamefont {Z\"ulicke}}, \bibinfo
  {author} {\bibfnamefont {J.}~\bibnamefont {Nitta}}, \ and\ \bibinfo {author}
  {\bibfnamefont {M.}~\bibnamefont {Kammermeier}},\ }\href {\doibase
  10.1103/PhysRevB.101.245417} {\bibfield  {journal} {\bibinfo  {journal}
  {Phys. Rev. B}\ }\textbf {\bibinfo {volume} {101}},\ \bibinfo {pages}
  {245417} (\bibinfo {year} {2020})}\BibitemShut {NoStop}%
\bibitem [{\citenamefont {Sawada}\ and\ \citenamefont
  {Koga}(2017)}]{Sawada2017}%
  \BibitemOpen
  \bibfield  {author} {\bibinfo {author} {\bibfnamefont {A.}~\bibnamefont
  {Sawada}}\ and\ \bibinfo {author} {\bibfnamefont {T.}~\bibnamefont {Koga}},\
  }\href {\doibase 10.1103/PhysRevE.95.023309} {\bibfield  {journal} {\bibinfo
  {journal} {Phys. Rev. E}\ }\textbf {\bibinfo {volume} {95}},\ \bibinfo
  {pages} {023309} (\bibinfo {year} {2017})}\BibitemShut {NoStop}%
\bibitem [{\citenamefont {Saito}\ \emph {et~al.}()\citenamefont {Saito},
  \citenamefont {Nishimura}, \citenamefont {Yoon}, \citenamefont {Kölzer},
  \citenamefont {Iizasa}, \citenamefont {Kammermeier}, \citenamefont
  {\mbox{Th.} {Sch\"apers}}, \citenamefont {Nitta},\ and\ \citenamefont
  {Kohda}}]{Saito2021}%
  \BibitemOpen
  \bibfield  {author} {\bibinfo {author} {\bibfnamefont {T.}~\bibnamefont
  {Saito}}, \bibinfo {author} {\bibfnamefont {T.}~\bibnamefont {Nishimura}},
  \bibinfo {author} {\bibfnamefont {J.}~\bibnamefont {Yoon}}, \bibinfo {author}
  {\bibfnamefont {J.}~\bibnamefont {Kölzer}}, \bibinfo {author} {\bibfnamefont
  {D.}~\bibnamefont {Iizasa}}, \bibinfo {author} {\bibfnamefont
  {M.}~\bibnamefont {Kammermeier}}, \bibinfo {author} {\bibnamefont {\mbox{Th.}
  {Sch\"apers}}}, \bibinfo {author} {\bibfnamefont {J.}~\bibnamefont {Nitta}},
  \ and\ \bibinfo {author} {\bibfnamefont {M.}~\bibnamefont {Kohda}},\
  }\href@noop {} {\ }\bibinfo {note} {(unpublished)}\BibitemShut {NoStop}%
\bibitem [{\citenamefont {Mal'shukov}\ and\ \citenamefont
  {Chao}(2000)}]{Malshukov2000}%
  \BibitemOpen
  \bibfield  {author} {\bibinfo {author} {\bibfnamefont {A.~G.}\ \bibnamefont
  {Mal'shukov}}\ and\ \bibinfo {author} {\bibfnamefont {K.~A.}\ \bibnamefont
  {Chao}},\ }\href {\doibase 10.1103/PhysRevB.61.R2413} {\bibfield  {journal}
  {\bibinfo  {journal} {Phys. Rev. B}\ }\textbf {\bibinfo {volume} {61}},\
  \bibinfo {pages} {R2413} (\bibinfo {year} {2000})}\BibitemShut {NoStop}%
\bibitem [{\citenamefont {Schwab}\ \emph {et~al.}(2006)\citenamefont {Schwab},
  \citenamefont {Dzierzawa}, \citenamefont {Gorini},\ and\ \citenamefont
  {Raimondi}}]{Schwab2006}%
  \BibitemOpen
  \bibfield  {author} {\bibinfo {author} {\bibfnamefont {P.}~\bibnamefont
  {Schwab}}, \bibinfo {author} {\bibfnamefont {M.}~\bibnamefont {Dzierzawa}},
  \bibinfo {author} {\bibfnamefont {C.}~\bibnamefont {Gorini}}, \ and\ \bibinfo
  {author} {\bibfnamefont {R.}~\bibnamefont {Raimondi}},\ }\href {\doibase
  10.1103/PhysRevB.74.155316} {\bibfield  {journal} {\bibinfo  {journal} {Phys.
  Rev. B}\ }\textbf {\bibinfo {volume} {74}},\ \bibinfo {pages} {155316}
  (\bibinfo {year} {2006})}\BibitemShut {NoStop}%
\bibitem [{\citenamefont {Yang}\ \emph {et~al.}(2010)\citenamefont {Yang},
  \citenamefont {Yan},\ and\ \citenamefont {Fardy}}]{Yang2010}%
  \BibitemOpen
  \bibfield  {author} {\bibinfo {author} {\bibfnamefont {P.}~\bibnamefont
  {Yang}}, \bibinfo {author} {\bibfnamefont {R.}~\bibnamefont {Yan}}, \ and\
  \bibinfo {author} {\bibfnamefont {M.}~\bibnamefont {Fardy}},\ }\href
  {\doibase 10.1021/nl100665r} {\bibfield  {journal} {\bibinfo  {journal} {Nano
  Lett.}\ }\textbf {\bibinfo {volume} {10}},\ \bibinfo {pages} {1529} (\bibinfo
  {year} {2010})}\BibitemShut {NoStop}%
\bibitem [{\citenamefont {L\"uffe}\ \emph {et~al.}(2011)\citenamefont
  {L\"uffe}, \citenamefont {Kailasvuori},\ and\ \citenamefont
  {Nunner}}]{Luffe2011}%
  \BibitemOpen
  \bibfield  {author} {\bibinfo {author} {\bibfnamefont {M.~C.}\ \bibnamefont
  {L\"uffe}}, \bibinfo {author} {\bibfnamefont {J.}~\bibnamefont
  {Kailasvuori}}, \ and\ \bibinfo {author} {\bibfnamefont {T.~S.}\ \bibnamefont
  {Nunner}},\ }\href {\doibase 10.1103/PhysRevB.84.075326} {\bibfield
  {journal} {\bibinfo  {journal} {Phys. Rev. B}\ }\textbf {\bibinfo {volume}
  {84}},\ \bibinfo {pages} {075326} (\bibinfo {year} {2011})}\BibitemShut
  {NoStop}%
\bibitem [{\citenamefont {Burkov}\ and\ \citenamefont
  {Balents}(2004)}]{Burkov2004}%
  \BibitemOpen
  \bibfield  {author} {\bibinfo {author} {\bibfnamefont {A.~A.}\ \bibnamefont
  {Burkov}}\ and\ \bibinfo {author} {\bibfnamefont {L.}~\bibnamefont
  {Balents}},\ }\href {\doibase 10.1103/PhysRevB.69.245312} {\bibfield
  {journal} {\bibinfo  {journal} {Phys. Rev. B}\ }\textbf {\bibinfo {volume}
  {69}},\ \bibinfo {pages} {245312} (\bibinfo {year} {2004})}\BibitemShut
  {NoStop}%
\bibitem [{\citenamefont {Stanescu}\ and\ \citenamefont
  {Galitski}(2007)}]{Stanescu2007}%
  \BibitemOpen
  \bibfield  {author} {\bibinfo {author} {\bibfnamefont {T.~D.}\ \bibnamefont
  {Stanescu}}\ and\ \bibinfo {author} {\bibfnamefont {V.}~\bibnamefont
  {Galitski}},\ }\href {\doibase 10.1103/PhysRevB.75.125307} {\bibfield
  {journal} {\bibinfo  {journal} {Phys. Rev. B}\ }\textbf {\bibinfo {volume}
  {75}},\ \bibinfo {pages} {125307} (\bibinfo {year} {2007})}\BibitemShut
  {NoStop}%
\bibitem [{\citenamefont {Wenk}\ and\ \citenamefont
  {Kettemann}(2010{\natexlab{a}})}]{Wenk2010}%
  \BibitemOpen
  \bibfield  {author} {\bibinfo {author} {\bibfnamefont {P.}~\bibnamefont
  {Wenk}}\ and\ \bibinfo {author} {\bibfnamefont {S.}~\bibnamefont
  {Kettemann}},\ }\href {\doibase 10.1103/PhysRevB.81.125309} {\bibfield
  {journal} {\bibinfo  {journal} {Phys. Rev. B}\ }\textbf {\bibinfo {volume}
  {81}},\ \bibinfo {pages} {125309} (\bibinfo {year}
  {2010}{\natexlab{a}})}\BibitemShut {NoStop}%
\bibitem [{\citenamefont {Liu}\ \emph {et~al.}(2012)\citenamefont {Liu},
  \citenamefont {Bundesmann},\ and\ \citenamefont {Richter}}]{Liu2012}%
  \BibitemOpen
  \bibfield  {author} {\bibinfo {author} {\bibfnamefont {M.-H.}\ \bibnamefont
  {Liu}}, \bibinfo {author} {\bibfnamefont {J.}~\bibnamefont {Bundesmann}}, \
  and\ \bibinfo {author} {\bibfnamefont {K.}~\bibnamefont {Richter}},\ }\href
  {\doibase 10.1103/PhysRevB.85.085406} {\bibfield  {journal} {\bibinfo
  {journal} {Phys. Rev. B}\ }\textbf {\bibinfo {volume} {85}},\ \bibinfo
  {pages} {085406} (\bibinfo {year} {2012})}\BibitemShut {NoStop}%
\bibitem [{\citenamefont {Wenk}\ and\ \citenamefont
  {Kettemann}(2010{\natexlab{b}})}]{wenkbook}%
  \BibitemOpen
  \bibfield  {author} {\bibinfo {author} {\bibfnamefont {P.}~\bibnamefont
  {Wenk}}\ and\ \bibinfo {author} {\bibfnamefont {S.}~\bibnamefont
  {Kettemann}},\ }\enquote {\bibinfo {title} {{Spin Relaxation in Quantum
  Wires}},}\ in\ \href {\doibase 10.1201/9781420075434} {\emph {\bibinfo
  {booktitle} {Handbook on Nanophysics}}},\ \bibinfo {editor} {edited by\
  \bibinfo {editor} {\bibfnamefont {K.}~\bibnamefont {Sattler}}}\ (\bibinfo
  {publisher} {Taylor \& Francis, Boca Raton},\ \bibinfo {year} {2010})\
  p.~\bibinfo {pages} {49}\BibitemShut {NoStop}%
\bibitem [{\citenamefont {Akkermans}\ and\ \citenamefont
  {Montambaux}(2017)}]{AkkermansBook}%
  \BibitemOpen
  \bibfield  {author} {\bibinfo {author} {\bibfnamefont {E.}~\bibnamefont
  {Akkermans}}\ and\ \bibinfo {author} {\bibfnamefont {G.}~\bibnamefont
  {Montambaux}},\ }\href@noop {} {\emph {\bibinfo {title} {Mesoscopic Physics
  of Electrons and Photons}}}\ (\bibinfo  {publisher} {Cambridge University
  Press},\ \bibinfo {year} {2017})\BibitemShut {NoStop}%
\bibitem [{\citenamefont {Kammermeier}(2018)}]{KammermeierPHD}%
  \BibitemOpen
  \bibfield  {author} {\bibinfo {author} {\bibfnamefont {M.}~\bibnamefont
  {Kammermeier}},\ }\emph {\bibinfo {title} {Control of Spin Relaxation in
  Disordered Quantum Wells and Nanowires}},\ \href
  {https://epub.uni-regensburg.de/37744/} {Ph.D. thesis},\ \bibinfo  {school}
  {University of Regensburg} (\bibinfo {year} {2018})\BibitemShut {NoStop}%
\bibitem [{Note1()}]{Note1}%
  \BibitemOpen
  \bibinfo {note} {Although $\protect \sqrt {D_e\tau }=l/\protect \sqrt {2}$ is
  smaller than the mean free path $l$, in the diffusive regime the exact value
  of the upper cut-off hardly affects the final result.}\BibitemShut {Stop}%
\bibitem [{\citenamefont {Beenakker}\ and\ \citenamefont {van
  Houten}(1991)}]{Beenakker1991}%
  \BibitemOpen
  \bibfield  {author} {\bibinfo {author} {\bibfnamefont {C.}~\bibnamefont
  {Beenakker}}\ and\ \bibinfo {author} {\bibfnamefont {H.}~\bibnamefont {van
  Houten}},\ }in\ \href {\doibase
  http://dx.doi.org/10.1016/S0081-1947(08)60091-0} {\emph {\bibinfo {booktitle}
  {Semiconductor Heterostructures and Nanostructures}}},\ \bibinfo {series}
  {Solid State Physics}, Vol.~\bibinfo {volume} {44},\ \bibinfo {editor}
  {edited by\ \bibinfo {editor} {\bibfnamefont {H.}~\bibnamefont {Ehrenreich}}\
  and\ \bibinfo {editor} {\bibfnamefont {D.}~\bibnamefont {Turnbull}}}\
  (\bibinfo  {publisher} {Academic Press},\ \bibinfo {year} {1991})\ pp.\
  \bibinfo {pages} {1 -- 228}\BibitemShut {NoStop}%
\bibitem [{\citenamefont {Bergmann}(1984)}]{Bergmann1984}%
  \BibitemOpen
  \bibfield  {author} {\bibinfo {author} {\bibfnamefont {G.}~\bibnamefont
  {Bergmann}},\ }\href {\doibase DOI: 10.1016/0370-1573(84)90103-0} {\bibfield
  {journal} {\bibinfo  {journal} {Phys. Rep.}\ }\textbf {\bibinfo {volume}
  {107}},\ \bibinfo {pages} {1} (\bibinfo {year} {1984})}\BibitemShut {NoStop}%
\bibitem [{\citenamefont {Dugaev}\ and\ \citenamefont
  {Khmel’nitskii}(1984)}]{Dugaev1984}%
  \BibitemOpen
  \bibfield  {author} {\bibinfo {author} {\bibfnamefont {V.~K.}\ \bibnamefont
  {Dugaev}}\ and\ \bibinfo {author} {\bibfnamefont {D.~E.}\ \bibnamefont
  {Khmel’nitskii}},\ }\href@noop {} {\bibfield  {journal} {\bibinfo
  {journal} {Zh. Eksp. Teor. Fiz.}\ }\textbf {\bibinfo {volume} {86}},\
  \bibinfo {pages} {1784} (\bibinfo {year} {1984})}\BibitemShut {NoStop}%
\bibitem [{\citenamefont {Meyer}\ \emph {et~al.}(2002)\citenamefont {Meyer},
  \citenamefont {Fal'ko},\ and\ \citenamefont {Altshuler}}]{Meyer2002}%
  \BibitemOpen
  \bibfield  {author} {\bibinfo {author} {\bibfnamefont {J.~S.}\ \bibnamefont
  {Meyer}}, \bibinfo {author} {\bibfnamefont {V.~I.}\ \bibnamefont {Fal'ko}}, \
  and\ \bibinfo {author} {\bibfnamefont {B.~L.}\ \bibnamefont {Altshuler}},\
  }in\ \href@noop {} {\emph {\bibinfo {booktitle} {Nato Science Series II}}},\
  Vol.~\bibinfo {volume} {72},\ \bibinfo {editor} {edited by\ \bibinfo {editor}
  {\bibfnamefont {I.~V.}\ \bibnamefont {Lerner}}, \bibinfo {editor}
  {\bibfnamefont {B.~L.}\ \bibnamefont {Altshuler}}, \bibinfo {editor}
  {\bibfnamefont {V.~I.}\ \bibnamefont {Fal'ko}}, \ and\ \bibinfo {editor}
  {\bibfnamefont {T.}~\bibnamefont {Giamarchi}}}\ (\bibinfo  {publisher}
  {Kluwer Academic Publishers, Dordrecht},\ \bibinfo {year} {2002})\ p.\
  \bibinfo {pages} {117}\BibitemShut {NoStop}%
\bibitem [{\citenamefont {Brouwer}\ \emph {et~al.}(2002)\citenamefont
  {Brouwer}, \citenamefont {Cremers},\ and\ \citenamefont
  {Halperin}}]{Brouwer2002}%
  \BibitemOpen
  \bibfield  {author} {\bibinfo {author} {\bibfnamefont {P.~W.}\ \bibnamefont
  {Brouwer}}, \bibinfo {author} {\bibfnamefont {J.~N. H.~J.}\ \bibnamefont
  {Cremers}}, \ and\ \bibinfo {author} {\bibfnamefont {B.~I.}\ \bibnamefont
  {Halperin}},\ }\href {\doibase 10.1103/PhysRevB.65.081302} {\bibfield
  {journal} {\bibinfo  {journal} {Phys. Rev. B}\ }\textbf {\bibinfo {volume}
  {65}},\ \bibinfo {pages} {081302(R)} (\bibinfo {year} {2002})}\BibitemShut
  {NoStop}%
\bibitem [{\citenamefont {Cremers}\ \emph {et~al.}(2003)\citenamefont
  {Cremers}, \citenamefont {Brouwer},\ and\ \citenamefont
  {Fal'ko}}]{Cremers2003}%
  \BibitemOpen
  \bibfield  {author} {\bibinfo {author} {\bibfnamefont {J.-H.}\ \bibnamefont
  {Cremers}}, \bibinfo {author} {\bibfnamefont {P.~W.}\ \bibnamefont
  {Brouwer}}, \ and\ \bibinfo {author} {\bibfnamefont {V.~I.}\ \bibnamefont
  {Fal'ko}},\ }\href {\doibase 10.1103/PhysRevB.68.125329} {\bibfield
  {journal} {\bibinfo  {journal} {Phys. Rev. B}\ }\textbf {\bibinfo {volume}
  {68}},\ \bibinfo {pages} {125329} (\bibinfo {year} {2003})}\BibitemShut
  {NoStop}%
\bibitem [{\citenamefont {Devries}\ and\ \citenamefont
  {Hasbrun}(1994)}]{DevriesBook1994}%
  \BibitemOpen
  \bibfield  {author} {\bibinfo {author} {\bibfnamefont {P.~L.}\ \bibnamefont
  {Devries}}\ and\ \bibinfo {author} {\bibfnamefont {J.~E.}\ \bibnamefont
  {Hasbrun}},\ }\href@noop {} {\emph {\bibinfo {title} {A first course in
  computational physics}}}\ (\bibinfo  {publisher} {Wiley},\ \bibinfo {year}
  {1994})\BibitemShut {NoStop}%
\bibitem [{\citenamefont {Sasaki}\ \emph {et~al.}(2014)\citenamefont {Sasaki},
  \citenamefont {Nonaka}, \citenamefont {Kunihashi}, \citenamefont {Kohda},
  \citenamefont {Bauernfeind}, \citenamefont {Dollinger}, \citenamefont
  {Richter},\ and\ \citenamefont {Nitta}}]{Sasaki2014}%
  \BibitemOpen
  \bibfield  {author} {\bibinfo {author} {\bibfnamefont {A.}~\bibnamefont
  {Sasaki}}, \bibinfo {author} {\bibfnamefont {S.}~\bibnamefont {Nonaka}},
  \bibinfo {author} {\bibfnamefont {Y.}~\bibnamefont {Kunihashi}}, \bibinfo
  {author} {\bibfnamefont {M.}~\bibnamefont {Kohda}}, \bibinfo {author}
  {\bibfnamefont {T.}~\bibnamefont {Bauernfeind}}, \bibinfo {author}
  {\bibfnamefont {T.}~\bibnamefont {Dollinger}}, \bibinfo {author}
  {\bibfnamefont {K.}~\bibnamefont {Richter}}, \ and\ \bibinfo {author}
  {\bibfnamefont {J.}~\bibnamefont {Nitta}},\ }\href {\doibase
  10.1038/nnano.2014.128} {\bibfield  {journal} {\bibinfo  {journal} {Nat.
  Nanotechnol.}\ }\textbf {\bibinfo {volume} {9}},\ \bibinfo {pages} {703}
  (\bibinfo {year} {2014})}\BibitemShut {NoStop}%
\bibitem [{\citenamefont {Nishimura}\ \emph {et~al.}(2021)\citenamefont
  {Nishimura}, \citenamefont {Yoshizumi}, \citenamefont {Saito}, \citenamefont
  {Iizasa}, \citenamefont {Nitta},\ and\ \citenamefont
  {Kohda}}]{Nishimura2021}%
  \BibitemOpen
  \bibfield  {author} {\bibinfo {author} {\bibfnamefont {T.}~\bibnamefont
  {Nishimura}}, \bibinfo {author} {\bibfnamefont {K.}~\bibnamefont
  {Yoshizumi}}, \bibinfo {author} {\bibfnamefont {T.}~\bibnamefont {Saito}},
  \bibinfo {author} {\bibfnamefont {D.}~\bibnamefont {Iizasa}}, \bibinfo
  {author} {\bibfnamefont {J.}~\bibnamefont {Nitta}}, \ and\ \bibinfo {author}
  {\bibfnamefont {M.}~\bibnamefont {Kohda}},\ }\href {\doibase
  10.1103/PhysRevB.103.094412} {\bibfield  {journal} {\bibinfo  {journal}
  {Phys. Rev. B}\ }\textbf {\bibinfo {volume} {103}},\ \bibinfo {pages}
  {094412} (\bibinfo {year} {2021})}\BibitemShut {NoStop}%
\bibitem [{Note2()}]{Note2}%
  \BibitemOpen
  \bibinfo {note} {To confirm the correctness of our numerical calculations, we
  reproduced Figs.~1 and 2 of Pikus \protect \textit {et al.} in Ref.~\protect
  \rev@citealpnum {Pikus1995}, who performed analogous numerical calculations
  but used the extension of the summation $N_{\protect \rm max }\rightarrow
  \infty $ equivalently to Iordanski \protect \textit {et al.}'s approach to
  smoothen their curves (cf. Sec.~\ref {sec:Iordanski}).}\BibitemShut {Stop}%
\bibitem [{Note3()}]{Note3}%
  \BibitemOpen
  \bibinfo {note} {Note that Ref.~\protect \rev@citealpnum {Saito2021} defines
  the $k$-linear and $k$-cubic Dresselhaus coefficient as $\beta _1$ and $\beta
  _3$, respectively, which are related by $\beta ^{(1)}=\beta _1-\beta _3$ and
  $\beta ^{(3)}=\beta _3$ to the definitions in the present paper.}\BibitemShut
  {Stop}%
\bibitem [{Note4()}]{Note4}%
  \BibitemOpen
  \bibinfo {note} {To ensure the correct implementation of Marinescu \protect
  \textit {et al.}'s formula, we numerically reproduced Fig.~2 in Ref.~\protect
  \rev@citealpnum {Marinescu2019}.}\BibitemShut {Stop}%
\bibitem [{\citenamefont {Ganichev}\ \emph {et~al.}(2002)\citenamefont
  {Ganichev}, \citenamefont {Danilov}, \citenamefont {Bel'kov}, \citenamefont
  {Ivchenko}, \citenamefont {Bichler}, \citenamefont {Wegscheider},
  \citenamefont {Weiss},\ and\ \citenamefont {Prettl}}]{Ganichev2002}%
  \BibitemOpen
  \bibfield  {author} {\bibinfo {author} {\bibfnamefont {S.~D.}\ \bibnamefont
  {Ganichev}}, \bibinfo {author} {\bibfnamefont {S.~N.}\ \bibnamefont
  {Danilov}}, \bibinfo {author} {\bibfnamefont {V.~V.}\ \bibnamefont
  {Bel'kov}}, \bibinfo {author} {\bibfnamefont {E.~L.}\ \bibnamefont
  {Ivchenko}}, \bibinfo {author} {\bibfnamefont {M.}~\bibnamefont {Bichler}},
  \bibinfo {author} {\bibfnamefont {W.}~\bibnamefont {Wegscheider}}, \bibinfo
  {author} {\bibfnamefont {D.}~\bibnamefont {Weiss}}, \ and\ \bibinfo {author}
  {\bibfnamefont {W.}~\bibnamefont {Prettl}},\ }\href {\doibase
  10.1103/PhysRevLett.88.057401} {\bibfield  {journal} {\bibinfo  {journal}
  {Phys. Rev. Lett.}\ }\textbf {\bibinfo {volume} {88}},\ \bibinfo {pages}
  {057401} (\bibinfo {year} {2002})}\BibitemShut {NoStop}%
\bibitem [{\citenamefont {Schneider}\ \emph {et~al.}(2004)\citenamefont
  {Schneider}, \citenamefont {Kainz}, \citenamefont {Ganichev}, \citenamefont
  {Danilov}, \citenamefont {Rössler}, \citenamefont {Wegscheider},
  \citenamefont {Weiss}, \citenamefont {Prettl}, \citenamefont {Bel’kov},
  \citenamefont {Glazov}, \citenamefont {Golub},\ and\ \citenamefont
  {Schuh}}]{Schneider2004}%
  \BibitemOpen
  \bibfield  {author} {\bibinfo {author} {\bibfnamefont {P.}~\bibnamefont
  {Schneider}}, \bibinfo {author} {\bibfnamefont {J.}~\bibnamefont {Kainz}},
  \bibinfo {author} {\bibfnamefont {S.~D.}\ \bibnamefont {Ganichev}}, \bibinfo
  {author} {\bibfnamefont {S.~N.}\ \bibnamefont {Danilov}}, \bibinfo {author}
  {\bibfnamefont {U.}~\bibnamefont {Rössler}}, \bibinfo {author}
  {\bibfnamefont {W.}~\bibnamefont {Wegscheider}}, \bibinfo {author}
  {\bibfnamefont {D.}~\bibnamefont {Weiss}}, \bibinfo {author} {\bibfnamefont
  {W.}~\bibnamefont {Prettl}}, \bibinfo {author} {\bibfnamefont {V.~V.}\
  \bibnamefont {Bel’kov}}, \bibinfo {author} {\bibfnamefont {M.~M.}\
  \bibnamefont {Glazov}}, \bibinfo {author} {\bibfnamefont {L.~E.}\
  \bibnamefont {Golub}}, \ and\ \bibinfo {author} {\bibfnamefont
  {D.}~\bibnamefont {Schuh}},\ }\href {\doibase 10.1063/1.1753656} {\bibfield
  {journal} {\bibinfo  {journal} {J. Appl. Phys.}\ }\textbf {\bibinfo {volume}
  {96}},\ \bibinfo {pages} {420} (\bibinfo {year} {2004})}\BibitemShut
  {NoStop}%
\bibitem [{\citenamefont {Gradl}\ \emph {et~al.}(2014)\citenamefont {Gradl},
  \citenamefont {Kempf}, \citenamefont {Schuh}, \citenamefont {Bougeard},
  \citenamefont {Winkler}, \citenamefont {Sch\"uller},\ and\ \citenamefont
  {Korn}}]{Gradl2014}%
  \BibitemOpen
  \bibfield  {author} {\bibinfo {author} {\bibfnamefont {C.}~\bibnamefont
  {Gradl}}, \bibinfo {author} {\bibfnamefont {M.}~\bibnamefont {Kempf}},
  \bibinfo {author} {\bibfnamefont {D.}~\bibnamefont {Schuh}}, \bibinfo
  {author} {\bibfnamefont {D.}~\bibnamefont {Bougeard}}, \bibinfo {author}
  {\bibfnamefont {R.}~\bibnamefont {Winkler}}, \bibinfo {author} {\bibfnamefont
  {C.}~\bibnamefont {Sch\"uller}}, \ and\ \bibinfo {author} {\bibfnamefont
  {T.}~\bibnamefont {Korn}},\ }\href {\doibase 10.1103/PhysRevB.90.165439}
  {\bibfield  {journal} {\bibinfo  {journal} {Phys. Rev. B}\ }\textbf {\bibinfo
  {volume} {90}},\ \bibinfo {pages} {165439} (\bibinfo {year}
  {2014})}\BibitemShut {NoStop}%
\bibitem [{\citenamefont {Gradl}\ \emph {et~al.}(2018)\citenamefont {Gradl},
  \citenamefont {Winkler}, \citenamefont {Kempf}, \citenamefont {Holler},
  \citenamefont {Schuh}, \citenamefont {Bougeard}, \citenamefont
  {Hern\'andez-M\'{\i}nguez}, \citenamefont {Biermann}, \citenamefont {Santos},
  \citenamefont {Sch\"uller},\ and\ \citenamefont {Korn}}]{Gradl2018}%
  \BibitemOpen
  \bibfield  {author} {\bibinfo {author} {\bibfnamefont {C.}~\bibnamefont
  {Gradl}}, \bibinfo {author} {\bibfnamefont {R.}~\bibnamefont {Winkler}},
  \bibinfo {author} {\bibfnamefont {M.}~\bibnamefont {Kempf}}, \bibinfo
  {author} {\bibfnamefont {J.}~\bibnamefont {Holler}}, \bibinfo {author}
  {\bibfnamefont {D.}~\bibnamefont {Schuh}}, \bibinfo {author} {\bibfnamefont
  {D.}~\bibnamefont {Bougeard}}, \bibinfo {author} {\bibfnamefont
  {A.}~\bibnamefont {Hern\'andez-M\'{\i}nguez}}, \bibinfo {author}
  {\bibfnamefont {K.}~\bibnamefont {Biermann}}, \bibinfo {author}
  {\bibfnamefont {P.~V.}\ \bibnamefont {Santos}}, \bibinfo {author}
  {\bibfnamefont {C.}~\bibnamefont {Sch\"uller}}, \ and\ \bibinfo {author}
  {\bibfnamefont {T.}~\bibnamefont {Korn}},\ }\href {\doibase
  10.1103/PhysRevX.8.021068} {\bibfield  {journal} {\bibinfo  {journal} {Phys.
  Rev. X}\ }\textbf {\bibinfo {volume} {8}},\ \bibinfo {pages} {021068}
  (\bibinfo {year} {2018})}\BibitemShut {NoStop}%
\bibitem [{\citenamefont {Ohno}\ \emph {et~al.}(1999)\citenamefont {Ohno},
  \citenamefont {Terauchi}, \citenamefont {Adachi}, \citenamefont {Matsukura},\
  and\ \citenamefont {Ohno}}]{Ohno1999}%
  \BibitemOpen
  \bibfield  {author} {\bibinfo {author} {\bibfnamefont {Y.}~\bibnamefont
  {Ohno}}, \bibinfo {author} {\bibfnamefont {R.}~\bibnamefont {Terauchi}},
  \bibinfo {author} {\bibfnamefont {T.}~\bibnamefont {Adachi}}, \bibinfo
  {author} {\bibfnamefont {F.}~\bibnamefont {Matsukura}}, \ and\ \bibinfo
  {author} {\bibfnamefont {H.}~\bibnamefont {Ohno}},\ }\href
  {http://link.aps.org/doi/10.1103/PhysRevLett.83.4196} {\bibfield  {journal}
  {\bibinfo  {journal} {Phys. Rev. Lett.}\ }\textbf {\bibinfo {volume} {83}},\
  \bibinfo {pages} {4196} (\bibinfo {year} {1999})}\BibitemShut {NoStop}%
\bibitem [{\citenamefont {M\"uller}\ \emph {et~al.}(2008)\citenamefont
  {M\"uller}, \citenamefont {R\"omer}, \citenamefont {Schuh}, \citenamefont
  {Wegscheider}, \citenamefont {H\"ubner},\ and\ \citenamefont
  {Oestreich}}]{Mueller2008}%
  \BibitemOpen
  \bibfield  {author} {\bibinfo {author} {\bibfnamefont {G.~M.}\ \bibnamefont
  {M\"uller}}, \bibinfo {author} {\bibfnamefont {M.}~\bibnamefont {R\"omer}},
  \bibinfo {author} {\bibfnamefont {D.}~\bibnamefont {Schuh}}, \bibinfo
  {author} {\bibfnamefont {W.}~\bibnamefont {Wegscheider}}, \bibinfo {author}
  {\bibfnamefont {J.}~\bibnamefont {H\"ubner}}, \ and\ \bibinfo {author}
  {\bibfnamefont {M.}~\bibnamefont {Oestreich}},\ }\href {\doibase
  10.1103/PhysRevLett.101.206601} {\bibfield  {journal} {\bibinfo  {journal}
  {Phys. Rev. Lett.}\ }\textbf {\bibinfo {volume} {101}},\ \bibinfo {pages}
  {206601} (\bibinfo {year} {2008})}\BibitemShut {NoStop}%
\bibitem [{\citenamefont {Chen}\ \emph {et~al.}(2014)\citenamefont {Chen},
  \citenamefont {F\"alt}, \citenamefont {Wegscheider},\ and\ \citenamefont
  {Salis}}]{Chen2014a}%
  \BibitemOpen
  \bibfield  {author} {\bibinfo {author} {\bibfnamefont {Y.~S.}\ \bibnamefont
  {Chen}}, \bibinfo {author} {\bibfnamefont {S.}~\bibnamefont {F\"alt}},
  \bibinfo {author} {\bibfnamefont {W.}~\bibnamefont {Wegscheider}}, \ and\
  \bibinfo {author} {\bibfnamefont {G.}~\bibnamefont {Salis}},\ }\href
  {\doibase 10.1103/PhysRevB.90.121304} {\bibfield  {journal} {\bibinfo
  {journal} {Phys. Rev. B}\ }\textbf {\bibinfo {volume} {90}},\ \bibinfo
  {pages} {121304(R)} (\bibinfo {year} {2014})}\BibitemShut {NoStop}%
\bibitem [{\citenamefont {Lampret}(2001)}]{EulerMacLaurinFormula}%
  \BibitemOpen
  \bibfield  {author} {\bibinfo {author} {\bibfnamefont {V.}~\bibnamefont
  {Lampret}},\ }\href {\doibase 10.1080/0025570X.2001.11953046} {\bibfield
  {journal} {\bibinfo  {journal} {Math. Mag.}\ }\textbf {\bibinfo {volume}
  {74}},\ \bibinfo {pages} {109} (\bibinfo {year} {2001})}\BibitemShut
  {NoStop}%
\end{thebibliography}%

\end{document}